\documentclass[a4paper,12pt]{article}
\usepackage[dvips]{graphicx}
\usepackage{amsmath,amsthm,amssymb,amscd,epsfig,latexsym}
\usepackage[T1]{fontenc}
\makeatother

\numberwithin{equation}{section}
\newcommand{\be}{\begin{equation}}
\newcommand{\ee}{\end{equation}}
\newcommand{\bea}{\begin{eqnarray}}
\newcommand{\eea}{\end{eqnarray}}

\newtheorem{thm}{Theorem}[section]
\newtheorem{lem}[thm]{Lemma}
\newtheorem{prop}[thm]{Proposition}
\newtheorem{cor}[thm]{Corollary}

\newtheorem{rmk}{Remark}[section]
\newtheorem{rem}[rmk]{Remark}

\font\twelvebb=msbm10 scaled 1100 \font\fivebb=msbm5 \font\sevenbb=msbm7
\newfam\bbfam  \scriptscriptfont\bbfam=\fivebb
\textfont\bbfam=\twelvebb \scriptfont\bbfam=\sevenbb

\def\N{{\mathbb N}}
\def\Z{{\mathbb Z}}
\def\R{{\mathbb R}}
\def\C{{\mathbb C}}

\def\d{\textrm{d}}
\def\one{{\mathchoice {\rm 1\mskip-4mu l} {\rm 1\mskip-4mu l} {\rm 1\mskip-4.5mu l} {\rm 1\mskip-5mu l}}}
\def\pf{\noindent{\bf Proof~:\ }}
\def\endproof{\hfill$\Box$\\}

\title{The ground state problem for a quantum Hamiltonian model describing friction}

\author{Laurent Bruneau \\ \\ 
Department of Mathematical Methods in Physics \\ 
Warsaw University \\ Ho\.{z}a 74, 00-682, Warszawa, Poland}

\begin{document}

\maketitle

\noindent \textbf{Abstract:} In this paper, we consider the quantum version of the hamiltonian model describing friction introduced in \cite{BDB}. This model consists of a particle which interacts with a bosonic reservoir representing a homogeneous medium through which the particle moves. We show that if the particle is confined, then the Hamiltonian admits a ground state if and only if a suitable infrared condition is
satisfied. The latter is violated in the case of linear friction, but satisfied when the
friction force is proportional to a higher power of the particle speed.

\section{Introduction}\label{sec:intro}

In \cite{BDB} we introduced a classical Hamiltonian model of a particle moving through a homogeneous dissipative medium at zero temperature in such a way that the particle experiences an effective \emph{linear} friction force proportional to its velocity. The medium consists at each point in the space of a vibration field with which the particle exchanges energy and momentum. More precisely the Hamiltonian is given by
\begin{eqnarray}\label{ham:class}
H(q,p,\phi,\pi) & = & \frac{p^2}{2}+V(q)+ \frac{1}{2}\int_{\R^d}dx \int_{\R^n} dy\, c^2|\nabla_y \phi(x,y)|^2+|\pi(x,y)|^2 \nonumber\\
 & & \qquad \qquad +\int_{\R^d}dx \int_{\R^n}dy\, \rho_1(x-q)\rho_2(y)\phi(x,y),
\end{eqnarray}
where $V$ is an external potential, $c$ represents the speed of the wave propagation in the ``membranes'' and the functions $\rho_1$ and $\rho_2$ determine the coupling between the particle and the field and are smooth radial functions with compact support. 

We studied the asymptotic behaviour of the particle motion for two categories of potentials: linear ones (which means constant external force) and confining ones. We proved that under suitable assumptions (on the initial conditions), for $c$ sufficiently large and, most importantly, $n=3,$ the particle behaves asymptotically as if its motion was governed by the effective equation
$$
\ddot{q}(t)+\gamma \dot{q}(t)= -\nabla V(q(t)),
$$
where the friction coefficient $\gamma$ is non negative and is explicit in terms of the parameters of the model:
\begin{equation}\label{eq:frictioncoef}
\gamma:= \frac{\pi}{c^3} |\hat{\rho}_2(0)|^2\int_{\R^n}d\xi \int_{\R^{d-1}}d\eta\, |\hat{\rho}_1(|\xi|,\eta)|^2.
\end{equation}
If $V=-F\cdot q,$ which means that we apply a constant external force $F$ to the particle, then this 
particle reaches exponentially fast (with rate $\gamma$) an asymptotic velocity $v(F)=\frac{F}{\gamma}$ 
which is proportional to the applied force (at least for small forces). This is, in particular, at the 
origin of Ohm's law. On the other hand, if $V$ is confining, the particle stops at one of the critical 
points of the potential, the convergence rate being still exponential (but with rate $\frac{\gamma}{2}$ as expected from the effective equation).

In \cite{BDB} we mostly concentrated on linear friction. This is why the $n=3$ assumption was required. 
However, for other values of $n \, (>3)$, our model still describes friction. Indeed, the reaction force of the environment on a particle moving with velocity $v$ takes the form $-\gamma |v|^{n-3}v$ (at least for small $v$ and where $\gamma$ is defined in (\ref{eq:frictioncoef})). One can therefore see that we have linear friction when $n=3,$ and otherwise a friction force which is proportional to some other power of the velocity of the particle. 

Such models, where a small system interacts with a \emph{large} environment, are called open systems. The reason for studying those models is usually to have a Hamiltonian description of dissipative phenomena. There exist several mechanisms leading to dissipation. Among them, two important, and very different, mechanisms are radiation damping and friction (which can be linear or not). As far as radiation damping is concerned, there exist many models, which are more or less related to electromagnetism. One example is the ``classical Nelson model''  
\begin{eqnarray*}
H_{\textrm{nels}}(q,p,\phi,\pi) & = & \frac{p^2}{2}+V(q)+ \frac{1}{2}\int_{\R^d}dx \left( |\nabla 
\phi(x)|^2+|\pi(x)|^2\right)\\
 & & \qquad \qquad \qquad \qquad \qquad \qquad +\int_{\R^d}dx \rho(x-q)\phi(x),
\end{eqnarray*}
which has been studied in \cite{KKS} (except for the kinetic energy of the particle which was $\sqrt{p^2+1}$ instead of $\frac{p^2}{2}$). This model describes a particle interacting with a scalar radiation field,
and exhibits radiation damping. Concerning friction, although there exist various Hamiltonian models in the literature, ours is the only one we are aware of that describes the friction produced by the motion of 
the particle through a homogeneous medium. In particular, the coupling between the medium and the 
particle is translationally invariant and hence non-linear in the particle variable. This means that no 
dipole approximation is assumed and is essential for a correct treatment of a constant external force
field. Despite the formal similarity between our model and the classical Nelson model, we want to stress 
once again that they describe physically totally different phenomena. This is reflected in mathematical 
differences that will become apparent below.

Our goal in this paper is to begin the study of the quantum version of the model (\ref{ham:class}). Since 
the speed of the wave propagation will not play any role in our paper, we take it equal to 1.
The quantum Hamiltonian can then be written as follows
\begin{eqnarray*}
H & = & \left( -\Delta +V\right) \otimes \one + \one \otimes \int dx\, dk\, \omega(x,k)a^*(x,k)a(x,k)\\
 & & \qquad \qquad \qquad \qquad +\int dx\, dk\, \frac{\rho_1(x-Q)\hat{\rho}_2(k)}{\sqrt{2\omega(x,k)}}\otimes a^*(x,k)+ h.c.,
\end{eqnarray*}
where $a$ and $a^*$ are the usual annihilation and creation operators on the bosonic Fock space 
$\mathcal{F}(L^2(\R^{d+n},dx\, dk)),$ and $\omega(x,k)=|k|$ is the bosons dispersion relation.  
In this paper, we start with the study of confining potentials, which are less difficult. More precisely,
we deal with the question of existence of a ground state, which is essential before studying questions such as scattering theory or return to equilibrium for example. If a Hamiltonian is bounded from below, we say that it admits a ground state if the infimum of its spectrum is an eigenvalue. We call ground state energy this infimum and ground state any corresponding eigenvector if it exists. We will prove that such a ground state exists provided the following \emph{infrared condition} is satisfied: 
$$
\int_{\R^n} dk\, \frac{|\hat{\rho}_2(k)|^2}{|k|^3} < +\infty.
$$
This condition will be used to control the number of bosons which have low energy (soft bosons). Let us suppose that $\hat{\rho}_2(0)\neq 0.$ Indeed, this is the only interesting case since the friction 
coefficient $\gamma$ vanishes together with $\hat{\rho}_2(0)$ (see (\ref{eq:frictioncoef})). Then, there exists a ground state if the infrared condition is satisfied (Theorem \ref{thm:gdstate}). One can see that this condition is fulfilled when the friction is non-linear. On the other hand, for linear friction, there is generically no ground state (Proposition \ref{prop:noground}). Thus, we have a class of models, depending on a parameter $n,$ describing friction phenomena, linear or proportional to a power of the velocity of the particle, for which we are able to say wether they admit a ground state or not.

As in the classical case, our model looks very similar to the Nelson model, and more generally to the Pauli-Fierz models (following the terminology of \cite{DG}), in which a (small) quantum system interacts with a scalar bosonic field, although they lead to very different dissipative phenomena. 
We will recall some basic facts about Fock spaces and describe the quantum version of the model in 
Sect. \ref{sec:model}, while, in Sect. \ref{sec:results}, we state our main results. 

To prove the existence of a ground sate, we follw the standard strategy: we first prove the result for coupling to a massive field and then we let the mass tend to zero. 
We study the massive case in Sect. \ref{sec:massifs} along the lines of \cite{BFS1}-\cite{BFS2}-\cite{GJ}: we first constrain the model to a finite box (|x|<L) and then control the error terms as $L$ goes to infinity. This has to be done with care since in the interaction term, the norm of $\rho_1(x-Q)$ as an operator on $L^2(\R^d)$ does not decrease with $x.$ In order to control this problem, we will need to use the exponential decay of the spectral projectors in the $Q$ variable. Furthermore, a second difficulty that arises is the following. The cutoff in space is equivalent to discretizing the model in the momentum variable. It is therefore equivalent to study the ``cutoff'' Hamiltonians and Hamiltonians ``discretized'' in momentum. In the case of the models for radiation damping, those discrete Hamiltonians are quite easy to study.
Indeed, the free discrete Hamiltonian has then purely discrete spectrum (the energy of the bosons can 
only take a discrete number of values), and, because the interaction is relatively bounded with respect 
to it, so has the full discrete Hamiltonian. Now, this will not be the case in our model because the 
energy of the bosons only depends on $k$ and not on the discretized momentum which comes from the fact 
that the dispersion relation $\omega$ only depends on $k.$ This is the main mathematical difference with the 
models for radiation damping. One then has to control the momentum of the bosons in the ``$x$-direction''.
A careful study of the discrete Hamiltonians will therefore be needed (Sect. \ref{ssec:disc}).

In Sect. \ref{sec:nonmassif} we first prove Theorem \ref{thm:gdstate}. To do this, we adapt the proof of \cite{G} to our model. In particular, we will need to control the momentum in ``$x$'' of the bosons. Moreover, we will also have to take into account that the norm of $\rho_1(x-Q)$ as an operator on $L^2(\R^d)$ does not depend on $x$ and is therefore not square integrable with respect to this variable. Once again, this will require to use the exponential decay of the spectral projectors in the $Q$ variable. Finally, we also prove Proposition \ref{prop:noground}.

We said that one of the main assumptions for the existence of a ground state in the massless case was
the so-called \emph{infrared condition} (see Sect. \ref{sec:results}). In Sect. \ref{sec:infrarouge}, we will present some classical interpretation of this infrared condition.


\section{Description of the model}\label{sec:model}

\subsection{Fock spaces}\label{ssec:fock}

In this section, we give a rather general (and brief) presentation of the different objects we will use in this paper. It will in particular allow us to fix notations. The reader will find a more detailed description in \emph{e.g.} \cite{DG}-\cite{RS2}.

Let $\mathfrak{h}$ be a complex Hilbert space, which is often called the $1-$particle space. Given $f,g$ in $\mathfrak{h},$ we denote by $\langle f;g\rangle$ their scalar product. It is chosen to be antilinear in the first variable and linear in the second variable. For $m \in \N,$ we define the $m-$particle sector as the $m$-fold symmetric tensor product of $\mathfrak{h}:\ \mathfrak{h}_m= \otimes^m_s \mathfrak{h},$
with $\mathfrak{h}_0= \C.$ 
We then define the Fock space over $\mathfrak{h}$ to be the direct sum 
$$
\mathcal{F}(\mathfrak{h}):= \oplus_{m=0}^{\infty} \mathfrak{h}_m.
$$
We will denote by $\Omega=(1,0,\dots)$ the vacuum vector and by $a^*$ and $a$ the usual 
creation/annihilation operators on $\mathcal{F}(\mathfrak{h})$ (\cite{RS2}, Chapter X.7).

In the case where $\mathfrak{h}= L^2(\R^{\nu}),$ we can rewrite those operators in the following way:
$$
a^*(h)=\int_{\R^{\nu}} dk\, h(k)a^*(k), \quad  a(h)=\int_{\R^{\nu}} dk\, \bar{h}(k)a(k),
$$
where $a^*(k)$ and $a(k)$ are the distributional creation and annihilation operators. 
They satisfy the usual canonical commutation relations:
\begin{equation} \label{ccr2}
[a(k),a^*(k')] = \delta(k-k'), \quad \left[a(k),a(k') \right] = [a^*(k),a^*(k')]=0.
\end{equation}
Finally, given an operator $b$ on $\mathfrak{h},$ we define:
\begin{eqnarray} \label{2quantif}
\mbox{d}\Gamma(b)&:& \mathcal{F}(\mathfrak{h}) \to \mathcal{F}(\mathfrak{h}) \nonumber \\ 
\mbox{d}\Gamma(b)|_{\mathfrak{h}_m}&:=& \sum_{j=1}^m \underbrace{\one\otimes \dots \otimes \one}_{j-1} \otimes b \otimes
\underbrace{\one \otimes \dots \otimes \one}_{m-j},
\end{eqnarray}
and 
\begin{equation} \label{2quantif2}
\Gamma(b):\ \mathcal{F}(\mathfrak{h}) \to \mathcal{F}(\mathfrak{h}), \quad  
\Gamma(b)|_{\mathfrak{h}_m}:= \ \underbrace{b\otimes \dots \otimes b}_m.
\end{equation}
The operator $\mbox{d}\Gamma(b)$ is called the second quantization of the operator $b.$ Note that when $b$ is selfadjoint, we have the following relation
$$
e^{i\textrm{d}\Gamma(b)}=\Gamma(e^{ib}).
$$
An operator which plays an important role is the number operator $N:=\mbox{d}\Gamma(\one).$


\subsection{Description of the model}\label{ssec:modq}

We can now introduce the quantum version of the model introduced in Sect. \ref{sec:intro}.  
The dynamics of the particle is given by the Schr\"odinger operator $H_p=-\Delta +V$ on $L^2(\R^d).$
Troughout this paper we will only consider confining potentials, so that $H_p$ has a compact resolvent and purely discrete spectrum.

The Hilbert space for the environment will be the bosonic Fock space over $L^2(\R^{d+n},dx\, dk).$ In what follows, we will just write
\begin{equation} \label{def:fieldspace}
\mathcal{F}:= \mathcal{F} (L^2(\R^{d+n}, dx\, dk)).
\end{equation}
The Hamiltonian of the field is given by
\begin{equation} \label{hamq:champ}
H_f:= \mbox{d}\Gamma(\omega),
\end{equation}
where $\omega$ is the multiplication operator on $L^2(\R^{d+n},dx\, dk)$ by the function
\begin{equation} \label{reldispers}
\omega: (x,k)\in \R^d\times \R^n\rightarrow \omega(x,k)=|k| \in [0,+\infty[.
\end{equation}
The function $\omega$ depends only on $k,$ so we will write $\omega(k)$ for $\omega(x,k).$
It is well known that one can rewrite $H_f$ using the creation and annihilation operators as follows:
\begin{equation} \label{hamq:champ2}
H_f= \int_{\R^{d+n}} dx \, dk \, \omega(k) a^*(x,k)a(x,k).
\end{equation}

We can now describe the full system. The Hilbert space is the tensor product of the particle space and of the environment one, namely:
\begin{equation} \label{def:space}
\mathcal{H}:= L^2(\R^d)\otimes \mathcal{F},
\end{equation}
and the free Hamiltonian ({\emph i.e.} without interaction) is given by:
\begin{equation} \label{hamq:libre}
H_0:=H_p \otimes \one + \one \otimes H_f.
\end{equation}

The interaction term is given by 
\begin{eqnarray}\label{hamq:interaction}
H_I & := & \int dx\, dk\,\,  \rho_1(x-Q)\frac{\hat{\rho}_2(k)}{\sqrt{2\omega(k)}}\otimes a^*(x,k) \nonumber \\
 & & \qquad \qquad \qquad + \rho_1(x-Q)\frac{\bar{\hat{\rho}}_2(k)}{\sqrt{2\omega(k)}}\otimes a(x,k),
\end{eqnarray}
where $\rho_1$ and $\rho_2$ are two smooth functions with compact support and spherical symetry, and $\rho_1(x-Q)$ is the multiplication operator on $L^2(\R^d)$ by the function $\rho_1(x-\cdot).$ 

Finally, the full Hamiltonian of the interacting system is therefore
\begin{equation} \label{hamq:total}
H:= H_0+H_I.
\end{equation}


\section{Main results}\label{sec:results}

\subsection{Selfadjointness}\label{ssec:aa}

From now, we will suppose that $n\geq 3.$ We first give the precise condition we impose on the potential $V:$
\begin{quote}
(C) $V \in L^2_{loc}(\R^d), \lim_{|q|\to \infty} V(q)=+\infty.$
\end{quote}
This hypothesis ensures that $H_p$ is well defined and is selfadjoint on  $\mathcal{D}(H_p)= 
\{\psi \in L^2(\R^d)| H_p\psi \in L^2(\R^d)\}$ (\cite{RS2}, Theorem X.28). We also know that $H_f$ is selfadjoint on its domain $\mathcal{D}(H_f)$ (\cite{RS1}, Chapter VIII.10). One then easily proves that $H_0$ is essentially selfadjoint on $\mathcal{D}(H_p)\otimes \mathcal{D}(H_f)$ (\cite{RS1}, Chapter VIII.10). 
We now have the following result
\begin{prop} \label{prop:aa}
Suppose that $n\geq 3,$ and $V$ satisfies condition (C). Then $H$ is selfadjoint on $\mathcal{D}(H)=\mathcal{D}(H_0)$. Moreover, $H$ is essentially selfadjoint on any core for $H_0,$ and it is bouded from below.
\end{prop}

This is in the standard way a consequence of the Kato-Rellich theorem (\cite{RS2}, Theorem X.12). The
only ingredient needed is that $H_I$ is infinitesimally $H_0$-bounded, which follows from the following 
lemma.
\begin{lem} \label{lem:a-estimates}
Under the hypothesis of Proposition \ref{prop:aa}, for all $\Psi \in \mathcal{D}(H_0)$, we have:
$$
(i) \qquad \| \int dx\, dk \frac{\hat{\rho}_2(k)}{\sqrt{\omega (k)}}\rho_1(x-Q)\otimes a(x,k) \Psi 
\|^2_{\mathcal{H}} \qquad \qquad \qquad \qquad \qquad 
$$
$$
\qquad \qquad \qquad \qquad \leq \Big[ \int dx\, dk |\rho_1(x)|^2 \frac{|\hat{\rho}_2(k)|^2}{\omega(k)^2}
\Big] \|(\one \otimes H_f^{\frac{1}{2}})\Psi \|^2_{\mathcal{H}}.
$$

$$
(ii) \qquad \| \int dx\, dk \frac{\hat{\rho}_2(k)}{\sqrt{\omega (k)}}\rho_1(x-Q)\otimes a^*(x,k) \Psi 
\|^2_{\mathcal{H}} \qquad \qquad \qquad \qquad \qquad 
$$
$$
\qquad \qquad \qquad \qquad \leq  \Big[ \int dx dk |\rho_1(x)|^2 \frac{|\hat{\rho}_2(k)|^2}{\omega(k)^2} \Big] 
 \|(\one \otimes H_f^{\frac{1}{2}})\Psi \|^2_{\mathcal{H}}
$$ 
$$ 
\qquad \qquad \qquad \qquad \qquad \qquad \qquad +\Big[ \int dx\, dk |\rho_1(x)|^2 \frac{|\hat{\rho}_2(k)|^2}{\omega(k)} \Big] \| \Psi\|^2_{\mathcal{H}}.
$$
\end{lem}

\begin{rem} (i) Such kind of estimates are well known \cite{A1}-\cite{BFS1}-\cite{DJ} and are sometimes called $N_{\tau}-$ estimates.

(ii) The $n \geq 3$ hypothesis ensures that the integrals on the right-hand side of both inequalities converge.
\end{rem}

\noindent{\bf Proof of Lemma \ref{lem:a-estimates}:} We use the fact that $\mathcal{H}$ is isomorphic to $L^2(\R^d, dq, \mathcal{F}).$ We then have:
\begin{eqnarray*}
 & & \| \int dx\, dk \frac{\hat{\rho}_2(k)}{\sqrt{\omega (k)}}\rho_1(x-Q)\otimes a(x,k)
\Psi \|^2_{\mathcal{H}}\\
 & = & \int_{\R^d} dq \| \int dx\, dk \frac{\hat{\rho}_2(k)}{\sqrt
{\omega (k)}}\rho_1(x-q) a(x,k)\Psi(q)\|^2_{\mathcal{F}}\\
 & = & \int_{\R^d} dq \| a(g_q)\Psi(q)\|^2_{\mathcal{F}},
\end{eqnarray*} 
where $g_q$ is the function 
$
g_q(x,k)= \frac{\hat{\rho}_2(k)}{\sqrt{\omega (k)}}\rho_1(x-q).
$
For all $q\in \R^d$, we have (see \cite{BFS1}):
\begin{eqnarray*}
\|a(g_q)\Psi (q)\|^2_{\mathcal{F}} & = & \| \int dx\, dk\, g_q(x,k) a(x,k)\Psi (q)\|^2_{\mathcal{F}}  \\
 & \leq & \left( \int dx\, dk\, \frac{|g_q(x,k)|}{\sqrt{\omega(k)}} \sqrt{\omega(k)} \|a(x,k) \Psi (q)\|
 \right)^2 \\
 & \leq & \Big[ \int dx\, dk \frac{|g_q(x,k)|^2}{\omega(k)} \Big] \int dx\, dk\, \omega(k) \|a(x,k) \Psi
 (q)\|^2. 
\end{eqnarray*}
But
$$
\int dx\, dk\, \omega(k) \|a(x,k) \Psi (q)\|^2= \langle \Psi (q); H_f \Psi (q)\rangle, 
$$
so,
\begin{eqnarray*}
\|a(g_q)\Psi (q)\|^2_{\mathcal{F}} & \leq & \Big[ \int dx\, dk \frac{|g_q(x,k)|^2}{\omega(k)} \Big] 
 \|H^{1/2}_f \Psi (q)\|^2_{\mathcal{F}}\\
 & \leq & \Big[ \int dx\, dk |\rho_1(x)|^2 \frac{|\hat{\rho}_2(k)|^2}
 {\omega(k)^2} \Big] \|H^{1/2}_f \Psi (q)\|^2_{\mathcal{F}}.
\end{eqnarray*}  
Finally,
\begin{eqnarray*}
 & & \| \int dx\, dk \frac{\hat{\rho}_2(k)}{\sqrt{\omega (k)}}\rho_1(x-Q)\otimes a(x,k)
\Psi \|^2_{\mathcal{H}}\\
 & \leq & \Big[ \int dx\, dk |\rho_1(x)|^2 \frac{|\hat{\rho}_2(k)|^2}
 {\omega(k)^2} \Big] \int_{\R^d} dy \|H^{1/2}_f \Psi (y)\|^2_{\mathcal{F}}\\
 & \leq & \Big[ \int dx\, dk |\rho_1(x)|^2 \frac{|\hat{\rho}_2(k)|^2}
 {\omega(k)^2} \Big] \| (\one \otimes H^{1/2}_f)\Psi \|^2_{\mathcal{H}},
\end{eqnarray*}
which proves $(i).$ One proves $(ii)$ in a similar way.
\endproof



\subsection{Existence of a ground state}\label{ssec:gdstate}

Let $E_0$ denote the ground state energy of $H.$ 
It is well known that one of the main obstacles to the existence of a ground state, in those models 
where a particle interacts with a field, comes from the so-called \emph{infrared catastroph}, which is 
due to the behaviour of $\omega(k)$ for small $k$ and in particular to the fact that $\omega(0)=0.$ We 
will then need the following condition on the coupling:
\begin{quote}
(IR) $\int_{\R^n} dk\, \frac{|\hat{\rho}_2(k)|^2}{\omega(k)^3}<+\infty.$
\end{quote}
This is what we call the infrared condition. We prove the following theorem, which is the main result of our paper: 
\begin{thm}\label{thm:gdstate} Suppose $n\geq 3,$  $V$ satisfies hypothesis (C), and $\hat{\rho}_2$ satisfies (IR). Then $H$ has a ground state.
\end{thm}

As we said in the introduction, this (IR) condition is satisfied when the friction is non-linear but not if it is linear. On the other way, in the case of the Nelson model, the same kind of condition is 
necessary and sufficient to have a ground state \cite{G}-\cite{LMS}. It is then reasonable to think this
is also true for our model. Indeed, we will prove that if the infrared condition is violated, then there
is no ground state but provided the following additional condition is satisfied
$$
\hat{\rho}_1(0)\neq 0,
$$ 
which means that the total charge of the particle does not vanish. More precisely, 
we prove the following result:
\begin{prop}\label{prop:noground} Suppose $n\geq 3,$ $V$ satisfies hypothesis (C), $\hat{\rho}_2$ does not satisfy (IR) and $\hat{\rho}_1(0)\neq 0,$ then $H$ has no ground state.
\end{prop}

To prove Theorem \ref{thm:gdstate}, we will need to study some ``intermediate'' models, and in particular 
to consider \emph{massive} bosons and to ``discretize'' space. The term \emph{massive} means that, 
instead of $\omega(k),$ we will consider a function $\omega_m(k)$ satisfying 
$$
(H_{\omega})\left\{ \begin{array}{l} \nabla \omega_m \in L^{\infty}(\R^n), \\ \lim_{|k|\to \infty} \omega_m(k)=+\infty,
\\ \inf \omega_m(k)=m>0. \end{array} \right.
$$
Our proof will use different methods developed in the literature \cite{BFS1}-\cite{BFS2}-\cite{DG}-\cite{G}-\cite{GJ}. 

Finally, we would like to emphasize that all the Hamiltonians we will deal with have the same structure 
as (\ref{hamq:total}) and so, a similar result to the one of Proposition \ref{prop:aa} is available for 
each of them.



\section{Ground state for massive bosons}\label{sec:massifs}

Our goal in this section is to prove a first result similar to Theorem \ref{thm:gdstate} but in the case 
of massive bosons (Theorem \ref{thm:massif}, Sect. \ref{ssec:massif}). We use the same approach as in 
\cite{GJ} and \cite{BFS1}. The idea is first to consider a finite box ($|x|<L$) and then to control the 
remaining part as $L$ goes to infinity. We will see, in Sect. \ref{ssec:massif}, that the ``cutoff'' model
so obtained can be written in the form (\ref{hdisc}). We therefore first study models of this latter type
(Theorem \ref{thm:disc}).

\subsection{Discrete models}\label{ssec:disc}

\subsubsection{Description}\label{sssec:description}

We consider Hamiltonians of te form
\begin{eqnarray}\label{hdisc}
H^{\d} & := & H_p \otimes \one + \one \otimes \sum_{l \in \Z^d} \int_{\R^n} dk\, \omega_m(k) a^*_l(k)a_l(k) \nonumber \\
 & & \qquad \qquad \qquad + \sum_{l \in \Z^d} \int_{\R^n} dk\, (\beta_l(k) \otimes a^*_l(k) + \bar{\beta}_l(k) \otimes a_l(k))\nonumber\\
 & = & H_0^{\d} + W^{\d}, 
\end{eqnarray}
on the space 
\begin{equation}\label{esp-disc1}
\mathcal{H}^{\d}:= L^2(\R^d)\otimes \mathcal{F}\left(l^2(\Z^d)\otimes L^2(\R^n)\right),
\end{equation}
and where the $\beta_l(k)$ satisfy \\

\begin{quote}
$(C_{\beta}) \quad \beta_l(k)= \zeta_l \frac{\hat{\rho}_2(k)}{\sqrt{2\omega_m(k)}}$ where $\zeta_l$ is a multiplication operator on $L^2(\R^d)$ such that $\sup_l \| |l|^s\zeta_l \| < +\infty$ for all $s>0$,\\
\end{quote}

\noindent $a_l(k)$ and $a^*_l(k)$ are the annihilation and creation operators on the space 
$\mathcal{F}\left(l^2(\Z^d)\otimes L^2(\R^n)\right),$ and for $l=(l_1,\dots, l_d)\in \Z^d,$ $|l|:=\sup_i |l_i|.$

We would like to stress that one can consider the Hamiltonians of the form (\ref{hdisc}) as models similar to ours, but with only a discrete set of ``membranes'' (situated at each $l \in \Z^d$) rather than a continuous one.

Now, let $E^{\d}_0$ denote the ground state energy for $H^{\d}.$ We will prove the following:
\begin{thm}\label{thm:disc} $\sigma_{ess}(H^{\d}) \subset \left[ E^{\d}_0+m, +\infty \right[.$ In particular, $H^{\d}$ has a ground state $\phi^{\d}_0$.
\end{thm}

\subsubsection{Cutoff models}\label{sssec:disc-cutoff}

In the following, $M$ will be a non negative number. On $\mathcal{H}^{\d},$ we define
\begin{eqnarray}\label{hdiscl1}
H^{\d}(M) & := & H_0^{\d} + \sum_{|l|\leq M} \int_{\R^n} dk\, (\beta_l(k) \otimes a^*_l(k) + \bar{\beta}_l(k) \otimes
a_l(k))\\
 & = & H_0^{\d} + W^{\d}(M). \nonumber
\end{eqnarray}
We also define
\begin{eqnarray}\label{hdiscl2}
\tilde{H}^{\d}(M) & := & H_p \otimes \one + \one \otimes \sum_{|l|\leq M} \int_{\R^n} dk\, \omega_m(k) a^*_l(k)a_l(k)+W^{\d}(M)\quad \\
 & = & \tilde{H}^{\d}_0(M) + W^{\d}(M), \nonumber
\end{eqnarray}
as an operator on the space 
\begin{equation}\label{esp-disc2}
\mathcal{H}^{\d}_M:= L^2(\R^d)\otimes \mathcal{F}\left(l^2(\Lambda_M)\otimes L^2(\R^n)\right),
\end{equation}
where $\Lambda_M=\{l \in \Z^d, |l|\leq M\},$ so that $l^2(\Lambda_M)$ is a finite dimensional space. Let $E^{\d}_0(M)$ (resp. $\tilde{E}^{\d}_0(M)$) be the ground state energy for $H^{\d}(M)$ (resp. $\tilde{H}^{\d}(M)$). Our goal is to get informations on $H^{\d}$ from the ones we will have on $H^{\d}(M)$ (taking the limit $M \to +\infty$). Thus, we first prove a result similar to Theorem \ref{thm:disc}, but for $H^{\d}(M).$

\begin{prop}\label{prop:sp-disc} $\sigma_{ess}(H^{\d}(M)) \subset \left[ E^{\d}_0(M)+m, +\infty \right[.$ In particular, $H^{\d}(M)$ has a ground state $\phi^{\d}_0(M).$ Moreover, $E^{\d}_0(M)=\tilde{E}^{\d}_0(M).$
\end{prop}

\begin{lem}\label{lem:sp-disc} $\sigma_{ess}(\tilde{H}^{\d}(M)) \subset \left[ \tilde{E}^{\d}_0(M)+m, +\infty \right[.$ In particular, $\tilde{H}^{\d}(M)$ has a ground state $\tilde{\phi}^{\d}_0(M).$
\end{lem}

\noindent{\bf Proof of Lemma \ref{lem:sp-disc}:} The set $\Lambda_M$ is finite. If its cardinal was one, we would have exactly the model studied in \cite{DG}, and the lemma would correspond to their Theorem 4.1. Having finitely many elements does not change anything and the result can be proven the same way.\endproof

\noindent{\bf Proof of Proposition \ref{prop:sp-disc}:} The proposition follows immediately from the preceding lemma using an identification between $\mathcal{H}^{\d}_M$ and some subspace of $\mathcal{H}^{\d},$ \cite{GJ}. Indeed, one can write
$$
l^2(\Z^d)\simeq l^2(\Lambda_M) \oplus l^2(\Lambda_M^c),
$$
$\Lambda_M^c$ denoting the complement of $\Lambda_M$ in $\Z^d,$ so one has
$$
\mathcal{F}\left(l^2(\Z^d) \otimes L^2(\R^n)\right) \simeq \mathcal{F}\left(l^2(\Lambda_M) \otimes L^2(\R^n)\right) \otimes
\mathcal{F}\left(l^2(\Lambda_M^c) \otimes L^2(\R^n)\right).
$$
And finally
$$
\mathcal{H}^{\d} \simeq \mathcal{H}^{\d}_M \otimes \mathcal{F}\left(l^2(\Lambda_M^c) \otimes L^2(\R^n)\right).
$$
One can then identify $\mathcal{H}^{\d}_M$ with $\mathcal{H}^{\d}_M \otimes \Omega^c_M$ where $\Omega^c_M$ is the vacuum of $\mathcal{F}\left(l^2(\Lambda_M^c) \otimes L^2(\R^n)\right).$ We can rewrite $\mathcal{H}^{\d}$ as
$$
\mathcal{H}^{\d} = \bigoplus_{j=0}^{+\infty} \left(\mathcal{H}^{\d}_M \otimes^j_s 
\left(l^2(\Lambda_L^c) \otimes L^2(\R^n)\right) \right)  =  \bigoplus_{j=0}^{+\infty} \mathcal{H}^{(j)}.
$$
Actually, we have 
$$
\mathcal{H}^{\d}_M=\mathcal{H}^{(0)} \quad \textrm{and} \quad (\mathcal{H}^{\d}_M)^{\perp}= \bigoplus_{j=1}^{+\infty} \mathcal{H}^{(j)}.
$$
One sees that the $\mathcal{H}^{(j)}$ are invariants for $H^{\d}(M).$ But, on $\mathcal{H}^{(j)},$ one has 
\begin{eqnarray*}
H^{\d}(M) & = & \tilde{H}^{\d}(M) \otimes \one + \one \otimes \sum_{|l|>L} \int_{\R^n} dk\, \omega_m(k) 
a^*_l(k)a_l(k) \\
 & \geq & \tilde{H}^{\d}(M) \otimes \one + mj,
\end{eqnarray*}
and on $\mathcal{H}^{(0)},$ 
$$
H^{\d}(M) = \tilde{H}^{\d}(M) \otimes \one.
$$
Then, we have
$$
\sigma \left( H^{\d}(M)|_{\mathcal{H}^{\d}_M}\right) =\sigma \left( \tilde{H}^{\d}(M)\right) \quad \textrm{and}
\quad \sigma_{ess} \left( H^{\d}(M)|_{\mathcal{H}^{\d}_M}\right) =\sigma_{ess} \left( \tilde{H}^{\d}(M)\right),
$$
and also
$$
\sigma_{ess} \left( H^{\d}(M)|_{(\mathcal{H}^{\d}_M)^{\perp}}\right) \subset \sigma \left(
H^{\d}(M)|_{(\mathcal{H}^{\d}_M)^{\perp}}\right) \subset \left[ \tilde{E}^{\d}_0(M)+m, +\infty \right[, 
$$
which ends the proof.
Moreover, one can remark that $\phi^{\d}_0(M)=\tilde{\phi}^{\d}_0(M)\otimes \Omega^c_M.$\endproof

\subsubsection{Removing the cutoff}\label{sssec:disc-remove}

We first prove some convergence results as $M$ goes to infinity.

\begin{prop}\label{prop:st-res-disc} $H^{\d}(M)$ converges to $H^{\d}$ in the strong resolvent sens.
\end{prop}

\pf We have
$$
H^{\d}-H^{\d}(M) =  W^{\d}-W^{\d}(M)
 = \sum_{|l|>M} \int_{\R^n} dk\, \beta_l(k)\otimes a^*_l(k)+ \bar{\beta}_l(k)\otimes a_l(k).
$$
Let $\psi \in D(H^{\d}_0).$ Using condition $(C_{\beta}),$ one has
\begin{eqnarray*}
\| \sum_{|l|>M} \int_{\R^n} dk\, \bar{\beta}_l(k)\otimes a_l(k) \psi \|  & \leq & \frac{C(s)}{1+M^s} 
\| \sum_{|l|>M} \int_{\R^n} dk\, \one \otimes a_l(k) \psi \| \\
 & \leq & \frac{C(s)}{1+M^s} \| (\one \otimes N^{\d})^{\frac{1}{2}} \psi \| \ \leq \ \frac{C(s)}{1+M^s} \| (H^{\d}_0)^{\frac{1}{2}} \psi \|.
\end{eqnarray*}
Then, using the commutation relations (\ref{ccr2}), we have 
\begin{eqnarray*}
\| \sum_{|l|>M} \int_{\R^n} dk\, \beta_l(k)\otimes a^*_l(k) \psi \|^2  & = & \| \sum_{|l|>M} \int_{\R^n} dk\,
\bar{\beta}_l(k)\otimes a_l(k) \psi \|^2 \\
 & & \qquad \qquad + \left( \sum_{|l|>M} \int_{\R^n} dk\, |\beta_l(k)|^2 \right) \| \psi \|^2.
\end{eqnarray*}
Finally, one gets
$$
\|H^{\d} \psi -H^{\d}(M) \psi \| \leq \frac{2C(s)}{1+M^s} \| (H^{\d}_0)^{\frac{1}{2}} \psi \| + \left( \sum_{|l|>M}
\int_{\R^n} dk\, |\beta_l(k)|^2 \right)^{\frac{1}{2}} \| \psi \|.
$$
Using condition $(C_{\beta}),$ one shows that the right hand side tends to zero as $M$ goes to infinity. So, $H^{\d}(M)$ converges strongly to $H^{\d}$ and then also in the strong resolvent sens (\cite{RS1}, Theorem VIII.25).\endproof

\begin{prop}\label{prop:vp-disc} $E^{\d}_0(M)$ is a decreasing function of $M$ which tends to $E^{\d}_0.$
\end{prop}

\pf We know that, if $\phi^{\d}_0(M)$ is a ground state for $H^{\d}(M),$ then 
$\phi^{\d}_0(M)=\tilde{\phi}^{\d}_0(M)\otimes \Omega^c_M,$ and so 
$$
\forall l\in \Lambda_M^c, \forall k\in \R^n, a_l(k) \phi^{\d}_0(M)=0.
$$
Let $M'>M,$
\begin{eqnarray*}
E^{\d}_0(M') & \leq & \langle \phi^{\d}_0(M); H^{\d}(M') \phi^{\d}_0(M) \rangle \\
 & \leq & \underbrace{\langle \phi^{\d}_0(M); H^{\d}(M) \phi^{\d}_0(M) \rangle}_{= E^{\d}_0(M)}+ \underbrace{\langle 
 \phi^{\d}_0(M); (W^{\d}(M')-W^{\d}(M)) \phi^{\d}_0(M) \rangle}_{=0}.
\end{eqnarray*}
So the function $E^{\d}_0(M)$ decreases. With the same argument, one proves that $E^{\d}_0(M)\geq E^{\d}_0.$ 
Then $E^{\d}_0(M)$ converges to some $E_{\infty}\geq E^{\d}_0.$ But $E^{d}_0\in \sigma (H^{\d})$ and 
$H^{\d}(M)$ converges to $H^{\d}$ in the strong resolvent sens, so (\cite{RS1}, Theorem VIII.24),
$$
\forall M>0, \exists E(M)\in \sigma (H^{\d}(M))/ E(M)\to E^{d}_0.
$$
Using the fact that $E^{\d}_0(M)$ is the ground state energy of $H^{\d}(M),$ we finally get $E_{\infty}= E^{\d}_0.$\endproof

\begin{prop}\label{prop:norm-cvg-disc} Let $\Delta$ be an interval bounded from above. For all $s>0,$ there exists $K(s,\Delta)>0$ such that
$$
\| \chi_{\Delta}(H^{\d})(W^{\d}-W^{\d}(M))\chi_{\Delta}(H^{\d})\| \leq \frac{K(s,\Delta)}{1+M^s}.
$$
\end{prop}

\pf Let $\phi, \psi \in \mathcal{H}^{\d}.$ We have 
\begin{eqnarray*}
 & & \left|\langle \phi; \chi_{\Delta}(H^{\d})(W^{\d}-W^{\d}(M))\chi_{\Delta}(H^{\d}) \psi \rangle \right| \\ 
 & = & |\langle \phi; \chi_{\Delta}(H^{\d}) (\sum_{|l|>M} \int_{\R^n} dk \, \beta_l(k)\otimes a^*_l(k) 
  +\bar{\beta}_l(k) \otimes a_l(k)) \chi_{\Delta}(H^{\d}) \psi \rangle | \\
 & \leq & |\langle \chi_{\Delta}(H^{\d}) \phi; (\sum_{|l|>M} \int_{\R^n} dk \, \bar{\beta}_l(k) \otimes
  a_l(k)) \chi_{\Delta}(H^{\d}) \psi \rangle | \\
 & & \qquad \qquad \qquad + |\langle (\sum_{|l|>M} \int_{\R^n} dk \, \bar{\beta}_l(k) \otimes a_l(k))\chi_{\Delta}(H^{\d}) 
  \phi; \chi_{\Delta}(H^{\d}) \psi \rangle |
\end{eqnarray*}
\begin{eqnarray*}
 & \leq & \| \phi \| \times \| (\sum_{|l|>M} \int_{\R^n}dk\, \bar{\beta}_l(k)\otimes a_l(k))\chi_{\Delta}(H^{\d})\psi \| \\
 & & \qquad \qquad \qquad \qquad + \| \psi \| \times \| (\sum_{|l|>M} \int_{\R^n}dk\, \bar{\beta}_l(k)\otimes a_l(k))\chi_{\Delta}(H^{\d})\phi \| \\
 & \leq & \frac{C(s)}{1+M^s} \left( \| \phi \| \times \| (\one \otimes N^{\d})^{\frac{1}{2}} \chi_{\Delta}(H^{\d}) \psi \| + \|
 \psi \| \times \| (\one \otimes N^{\d})^{\frac{1}{2}} \chi_{\Delta}(H^{\d}) \phi \| \right).
\end{eqnarray*}
But $\Delta$ is bounded from above, $\one \otimes N^{\d} \leq \frac{1}{m} H_0^{\d}$ and $W^{\d}$ is relatively $H_0^{\d}$ bounded, so $(\one \otimes N^{\d})^{\frac{1}{2}} \chi_{\Delta}(H^{\d})$ is a bounded operator. Finally, one has 
$$
\left|\langle \phi; \chi_{\Delta}(H^{\d})(W^{\d}-W^{\d}(M))\chi_{\Delta}(H^{\d}) \psi \rangle \right| \leq \frac{2C(s)
\|(N^{\d})^{\frac{1}{2}} \chi_{\Delta}(H^{\d})\|}{1+M^s} \| \phi \| \times \| \psi \|,
$$
which ends the proof.\endproof

\noindent{\bf Proof of Theorem \ref{thm:disc}:}

We use the method of \cite{BFS2}.
Given an operator $A,$ $[A]_-$ will denote its negative part and $\textrm{Tr}(A)$ its trace. 
To prove the theorem, it suffices to show that, for all $\epsilon>0,$ we have 
$$
\textrm{Tr} \left\{ [ H^{\d} -E^{\d}_0-m+\epsilon]_- \right\} >-\infty.
$$
Let $\epsilon >0,$ and $\Delta= ]-\infty, E^{\d}_0+m-\epsilon[.$ Then
$$
[ H^{\d} -E^{\d}_0-m+\epsilon]_- = \chi_{\Delta}(H^{\d}) ( H^{\d} -E^{\d}_0-m+\epsilon) \chi_{\Delta}(H^{\d}),
$$
and so
\begin{eqnarray*}
\textrm{Tr} \left\{ [ H^{\d} -E^{\d}_0-m+\epsilon]_- \right\}  & = & \textrm{Tr} \left\{ \chi_{\Delta}(H^{\d}) (H^{\d} 
-E^{\d}_0-m+\epsilon) \chi_{\Delta}(H^{\d}) \right\} \\
 & = & \textrm{Tr} \{ \chi_{\Delta}(H^{\d}) (H^{\d}(M)-E^{\d}_0(M)-m+\epsilon \\
 & & \qquad +W^{\d}-W^{\d}(M) -E^{\d}_0+E^{\d}_0(M)) \chi_{\Delta}(H^{\d}) \}.
\end{eqnarray*}
But 
$$
E^{\d}_0(M) \to E^{\d}_0 \quad \textrm{and} \quad \| \chi_{\Delta}(H^{\d})(W^{\d}-W^{\d}(M))\chi_{\Delta}(H^{\d})\| \to
0,
$$
using Propositions \ref{prop:vp-disc} and \ref{prop:norm-cvg-disc}, so, for $M$ large enough, we have 
$$
\begin{array}{l}
\textrm{Tr} \left\{ [ H^{\d} -E^{\d}_0-m+\epsilon]_- \right\}\\
\qquad \qquad \qquad \geq \textrm{Tr} \left\{ \chi_{\Delta}(H^{\d}) 
(H^{\d}(M) -E^{\d}_0(M)-m+\frac{\epsilon}{2}) \chi_{\Delta}(H^{\d}) \right\} \\
\qquad \qquad \qquad \geq \textrm{Tr} \left\{ \chi_{\Delta}(H^{\d}) [H^{\d}(M) -E^{\d}_0(M)-m+\frac{\epsilon}{2}]_- \chi_{\Delta}(H^{\d}) \right\} \\
\qquad \qquad \qquad \geq \textrm{Tr} \left\{ [H^{\d}(M) -E^{\d}_0(M)-m+\frac{\epsilon}{2}]_- \right\}>-\infty\\
\end{array}
$$
where in the last step we used Proposition \ref{prop:sp-disc}.\endproof


\subsection{Continuous models}\label{ssec:massif}

In this section, we are interested in the model introduced in Sect. \ref{ssec:modq} but for massive bosons, \emph{i.e.} the function $\omega(k)$ is replaced by $\omega_m(k)$ satisfying $(H_{\omega}).$ We then consider, on $\mathcal{H},$ the following Hamiltonian:
\begin{eqnarray}\label{h-mass}
H_m & := & H_p \otimes \one + \one \otimes \int_{\R^d} dx\, \int_{\R^n} dk\, \omega_m(k) a^*(x,k)a(x,k) \nonumber \\
 & & \qquad +\int_{\R^d} dx\, \int_{\R^n} dk\, \rho_1(x-Q)\frac{\hat{\rho}_2(k)}{\sqrt{2\omega_m(k)}} \otimes a^*(x,k) \nonumber \\  
 & & \qquad \qquad \qquad \qquad +\rho_1(x-Q)\frac{\bar{\hat{\rho}}_2(k)}{\sqrt{2\omega_m(k)}} \otimes a(x,k) \\
 & = & H^0_m+ W_m. \nonumber
\end{eqnarray}

We denote by $E_m$ the ground state energy of $H_m.$ The main result of this section is the
\begin{thm}\label{thm:massif} $\sigma_{ess}(H_m) \subset [E_m+m, +\infty[.$ In particular, $H_m$ has a ground state $\phi_m.$
\end{thm}
The strategy of the proof is very similar to the one of the previous section. However, one has to be more careful with the estimates when removing the cutoff because the norm of $\rho_1(x-Q)$ as an operator on $L^2(\R^d)$ does not decrease with $x,$ even worse, it does not depend on it. To control this problem, we will use the exponential decay of the spectral projectors in the $Q$ variable, which will be obtained via the Agmon method (see Sect. \ref{sssec:expbound})

\subsubsection{Cutoff models}\label{sssec:cont-cutoff}

Let $j$ be a smooth function with compact support on $\R^d$ such that
$$
0 \leq j(x) \leq 1, \quad j(x)=1 \ \ \rm{for} \ |x|\leq 1/2,  \quad \rm{and} \quad j(x)=0 \ \
\rm{for} \ |x|\geq 3/4.
$$
For all $L>0,$ we define $j_L(x)=j(\frac{x}{L})$ and $\bar{j}_L(x)=1-j_L(x).$ We then define 
\begin{eqnarray}\label{h-massl1}
H_m(L) & := & H_m^0+ \int_{\R^d} dx\, \int_{\R^n} dk\, \rho_1(x-Q)j_L(x)\frac{\hat{\rho}_2(k)}{\sqrt{2\omega_m(k)}}
  \otimes a^*(x,k) \nonumber \\
 & & \qquad \qquad \qquad +\rho_1(x-Q)j_L(x)\frac{\bar{\hat{\rho}}_2(k)}{\sqrt{2\omega_m(k)}} \otimes a(x,k)\\
 & = & H_m^0 + W_m(L) \nonumber
\end{eqnarray}
on $\mathcal{H}.$ Using the definition of $j_L,$ one can, in $W_m(L),$ replace $\int_{\R^d} dx$ by 
$\int_{[-L,L]^d} dx.$ Finally, we define
\begin{equation}\label{h-massl2}
\tilde{H}_m(L):= H_p\otimes \one+ \one \otimes \int_{[-L,L]^d} dx\, \int_{\R^n} dk\, \omega_m(k) a^*(x,k)a(x,k)+ W_m(L)
\end{equation}
on $L^2(\R^d)\otimes \mathcal{F}\left(L^2([-L,L]^d)\otimes L^2(\R^n)\right).$ We denote by $E_m(L)$ and
$\tilde{E}_m(L)$ the ground state energies of those two operators respectively.

We have ``cut'' the Hamiltonian $H_m$ in the $x$ variable. We are now in a finite volume box. If we consider the variable $p,$ conjugate to $x,$ this is equivalent to ``discretizing'' the problem. One has to note that here the variable $p$ is discrete:  $p\in \Z^d.$ If 
$$
a^*_p(k)= \frac{1}{(2L)^{\frac{d}{2}}}\int_{[-L,L]^d} dx \, e^{ipx} a^*(x,k),\quad
a_p(k)= \frac{1}{(2L)^{\frac{d}{2}}}\int_{[-L,L]^d} dx \, e^{-ipx} a(x,k)
$$ 
and
$$
\beta_p= \frac{1}{(2L)^{\frac{d}{2}}}\int_{[-L,L]^d} dx \, \rho_1(x-Q)j_L(x)
$$ 
denote the Fourier coefficients of $a^*(x,k), a(x,k)$ and $\rho_1(x-Q)j_L(x)$ respectively, the problem can now be written as follows 
\begin{eqnarray*}
\tilde{H}_m(L)& = & H_p\otimes \one+ \one \otimes \sum_{p \in \Z^d} \int_{\R^n} dk\, \omega_m(k) a^*_p(k)a_p(k)  \\
 & & \qquad + \sum_{p \in \Z^d} \int_{\R^n} dk\, (\beta_p \frac{\hat{\rho}_2(k)}{\sqrt{2\omega_m(k)}}\otimes a^*_p(k) 
  + \bar{\beta}_p \frac{\bar{\hat{\rho}}_2(k)}{\sqrt{2\omega_m(k)}}\otimes a_p(k)),
\end{eqnarray*}
which has the form (\ref{hdisc}). If the  $\beta_p$ satisfy $(C_{\beta}),$ we will then have the following result:
\begin{prop}\label{prop:sp-mass1} $\forall L>0, \sigma_{ess}(\tilde{H}_m(L)) \subset [\tilde{E}_m(L)+m, +\infty[.$
\end{prop}
\noindent Finally, a splitting of $L^2(\R^d)$ into $L^2([-L,L]^d)\oplus L^2(\R^d\backslash [-L,L]^d)$ together with the argument of the previous section will lead to the
\begin{prop}\label{prop:sp-mass2} $\sigma_{ess}(H_m(L)) \subset [E_m(L)+m, +\infty[.$ In particular, $H_m(L)$ has a ground state $\phi_m(L).$
\end{prop}
So, it remains to check that the $\beta_p$ satisfy the condition $(C_{\beta}).$ The function $j_L$ is zero for $|x|>L$ and $\rho_1$ has compact support (in a ball of radius $R_1$), so  
$$
\forall |q|>L+R_1, \forall x\in \R^d, \, \rho_1(x-q)j_L(x)=0.
$$  
Then, for all $p$ in $\Z^d,$ $\beta_p$ is a multiplication operator by a compactly supported function. Moreover, the function $\rho_1(x-q)j_L(x)$ is $C^{\infty},$ so its Fourier coefficients decay faster than any power of $p.$ Those two facts ensure us that
$\sup_p \sup_q \big|\beta_p(q)|p|^n\big|<C_n(L)<+\infty$ and so condition $(C_{\beta})$ is satisfied. To prove Theorem \ref{thm:massif}, it remains to control the limit $L\to +\infty.$

\subsubsection{Exponential bounds}\label{sssec:expbound}

\begin{prop}\label{prop:exp-decr} Let $\Delta$ be a bounded from above interval. For any $\alpha>0,$ there exists $M(\alpha,\Delta)>0$ such that 
\begin{itemize}
\item[-] $\| (e^{\alpha|Q|}\otimes \one) \chi_{\Delta}(H_m(L))\| \leq M(\alpha, \Delta).$ 
\item[-] $\| (e^{\alpha|Q|}\otimes \one) \chi_{\Delta}(H_m)\| \leq M(\alpha, \Delta).$
\item[-] $\| (e^{\alpha|Q|}\otimes \one) \chi_{\Delta}(H)\| \leq M(\alpha, \Delta).$
\end{itemize}
\end{prop}
\noindent This bound is uniform in $L$ and $m.$ The proof is exactly the same as the one of Theorem II.1 of 
\cite{BFS1}. The only difference is that $\sigma_{ess}(H_p)=\emptyset,$ which makes things easier and in particular one does not need any condition on $\alpha$ or on the supremum of the interval $\Delta.$

For any $R>0,$ we now define
\begin{equation}\label{nb-photproches}
N(|x|\leq R):= \int_{|x|\leq R}dx\, \int_{\R^n}dk\, a^*(x,k)a(x,k),
\end{equation}
and
\begin{equation}\label{nb-photloin}
N(|x|>R):= \int_{|x|>R}dx\, \int_{\R^n}dk\, a^*(x,k)a(x,k).
\end{equation}
$N(|x|\leq R)$ is the number of bosons inside the ball centered at the origin and of radius $R$ (in the $x$ variable), and $N(|x|>R)$ is the number of bosons outside this ball. We will prove that the number of these ``far away'' bosons decays exponentially fast with $R.$ More precisely, we have the following estimate:
\begin{prop}\label{prop:photdecr} For any $\alpha>0,$ there exists $C(\alpha)>0$ such that 
\begin{equation}\label{eq:photdecr}
\langle \phi_m(L); \one \otimes N(|x|>R) \phi_m(L)\rangle \leq C(\alpha) e^{-\alpha R}
\end{equation}
uniformly in $L.$
\end{prop}
The idea is to adapt the proof of \cite{BFS1}. What is new in our model is that we need an explicit control on the number of ``far away'' bosons in the $x$ direction, even for massive bosons. Fot that purpose, we use the following lemma which comes from the well known pullthrough formula (see \emph{e.g.} \cite{G}):
\begin{lem}\label{lem:pullthrough} $\|\one \otimes a(x,k) \phi_m(L)\| \leq \frac{1}{\omega_m(k)}\|
\rho_1(x-Q)j_L(x)\frac{\hat{\rho}_2(k)}{\sqrt{2\omega_m(k)}}\otimes \one \phi_m(L)\|.$
\end{lem}


\noindent{\bf Proof of Proposition \ref{prop:photdecr}~:\ } Let $\alpha >0,$ 
\begin{eqnarray*}
 & & \langle \phi_m(L); \one \otimes N(|x|>R) \phi_m(L)\rangle \\
 & = & \int_{|x|>R}dx\, \int_{\R^n} dk\, \| \one \otimes a(x,k) \phi_m(L)\|^2 \\
 & \leq & \int_{|x|>R}dx\, \int_{\R^n} dk\, \frac{1}{\omega_m^2(k)} \| \rho_1(x-Q)j_L(x)
  \frac{\hat{\rho}_2(k)}{\sqrt{2\omega_m(k)}}\otimes \one \phi_m(L)\|^2 \\
 & \leq & \int_{|x|>R}dx\, \int_{\R^n} dk\, \frac{|\hat{\rho}_2(k)|^2}{2\omega_m^3(k)} \| \rho_1(x-Q) j_L(x) 
 e^{-\alpha |Q|}\|^2_{\mathcal{B}(L^2)} \times \|e^{\alpha |Q|}\otimes \one \phi_m(L)\|^2.
\end{eqnarray*}

The function $\hat{\rho}_2$ is a Schwartz function and $\omega_m$ is bounded from below by $m>0,$ so the integral with respect to the $k$ variable converges. Now we recall that the function $\rho_1$ has compact support in the ball of radius $R_1,$ so, for any given $x\in \R^d,$ we have 
\begin{eqnarray*}
\|\rho_1(x-Q)e^{-\alpha |Q|}\|_{\mathcal{B}(L^2)} & = & \sup_{q\in \R^d} |\rho_1(x-q)e^{-\alpha |q|}| \\
 & = & \sup_{|q-x|\leq R_1} |\rho_1(x-q)e^{-\alpha |q|}| \;
 \leq \| \rho_1 \|_{\infty} e^{\alpha R_1} e^{-\alpha |x|}.
\end{eqnarray*}
Thus
$$
\int_{|x|>R} dx\, \|\rho_1(x-Q)e^{-\alpha |Q|}\|^2_{\mathcal{B}(L^2)}  \leq  \| \rho_1 \|^2_{\infty} e^{2\alpha
R_1} \int_{|x|>R} dx\,  e^{-2\alpha |x|}
  \leq  K(\alpha) e^{-\alpha R}.
$$
And so, finally,
$$
\langle \phi_m(L); \one \otimes N(|x|>R) \phi_m(L)\rangle \leq K'(\alpha)e^{-\alpha R} \|e^{\alpha |Q|}\otimes \one \phi_m(L)\|^2.
$$
 
But, for any $L,$ we have $E_m(L)\leq E^0_{p}$ where $E^0_{p}$ is the ground state energy of $H_p.$ Indeed, if $\psi^0_p$ is the ground state of $H_p,$ we have
$$
E_m(L) \leq \langle \psi^0_p \otimes \Omega ; H_m(L)\; \psi^0_p \otimes \Omega\rangle = E_p^0.
$$
Take now $\Delta= ]-\infty, E_p^0],$ one can then write $\phi_m(L)= \chi_{\Delta}(H_m(L)) \phi_m(L),$ and so
\begin{eqnarray*}
\|e^{\alpha |Q|}\otimes \one \phi_m(L)\|^2 & \leq & \|e^{\alpha |Q|}\otimes \one \; \chi_{\Delta}(H_m(L))\|^2 \|\phi_m(L)\| \\
 & \leq & M(\alpha, \Delta)^2,
\end{eqnarray*}
which ends the proof. \endproof

We finally give an estimate similar to the one of Proposition \ref{prop:norm-cvg-disc}.
\begin{prop}\label{prop:norm-cvg-mass} Let $\Delta$ and $\alpha$ be as in Proposition \ref{prop:exp-decr}, then there exists $K(\alpha, \Delta)$ such that 
$$
\|\chi_{\Delta}(H_m) (W_m-W_m(L))\chi_{\Delta}(H_m)\| \leq K(\alpha, \Delta) e^{-\alpha L}.
$$
\end{prop}

\pf We follow the scheme of the proof of Proposition \ref{prop:exp-decr} using estimates similar to the ones of the previous proposition. Let $\phi, \psi \in \mathcal{H},$
\begin{eqnarray*}
 & & |\langle \phi; \chi_{\Delta}(H_m) (W_m-W_m(L))\chi_{\Delta}(H_m)\psi \rangle | \\
 & \leq & |\langle (e^{2\alpha |Q|}\otimes \one) \chi_{\Delta}(H_m)\phi; \Big(\int_{|x|>\frac{L}{2}}dx\, \int_{\R^n}dk\, 
  e^{-2\alpha |Q|}\rho_1(x-Q)\bar{j}_L(x) \\ 
 & &  \qquad \qquad \qquad \qquad \qquad \qquad \qquad \times \frac{\bar{\hat{\rho}}_2(k)}{\sqrt{2\omega_m(k)}} \otimes
  a(x,k)\Big) \chi_{\Delta}(H_m)\psi \rangle |\\ 
 & & \qquad + |\langle (e^{2\alpha |Q|}\otimes \one) \chi_{\Delta}(H_m)\psi; \Big(\int_{|x|>\frac{L}{2}}dx\, \int_{\R^n}dk\, 
  e^{-2\alpha |Q|}\rho_1(x-Q)\bar{j}_L(x) \\ 
 & &  \qquad \qquad \qquad \qquad \qquad \qquad \quad \qquad \times \frac{\bar{\hat{\rho}}_2(k)}{\sqrt{2\omega_m(k)}} \otimes
  a(x,k)\Big) \chi_{\Delta}(H_m)\phi \rangle |. 
\end{eqnarray*}
We consider only the first term, the other one being similar.
\begin{eqnarray*}
 & & |\langle (e^{2\alpha |Q|}\otimes \one) \chi_{\Delta}(H_m)\phi; \Big( \int_{|x|>\frac{L}{2}}dx\, \int_{\R^n}dk\, 
  e^{-2\alpha |Q|}\rho_1(x-Q)\bar{j}_L(x) \\ 
 & &  \qquad \qquad \qquad \qquad \qquad \qquad \qquad \times \frac{\bar{\hat{\rho}}_2(k)}{\sqrt{2\omega_m(k)}} \otimes
  a(x,k)\Big) \chi_{\Delta}(H_m)\psi \rangle |\\
 & \leq & \|(e^{2\alpha |Q|}\otimes \one) \chi_{\Delta}(H_m)\phi\| \times \| (\int_{|x|>\frac{L}{2}}dx\, \int_{\R^n}dk\, 
  e^{-2\alpha |Q|}\rho_1(x-Q)\bar{j}_L(x) \\
 & & \qquad \qquad \qquad \qquad \qquad \qquad \qquad \quad \times \frac{\bar{\hat{\rho}}_2(k)}{\sqrt{2\omega_m(k)}} \otimes
  a(x,k))\chi_{\Delta}(H_m)\psi\| \\
 & \leq & M(2\alpha, \Delta) \|\phi \| \left[ \int_{|x|>\frac{L}{2}}dx\, \int_{\R^n}dk\, \| e^{-2\alpha |Q|}\rho_1(x-Q)
  \bar{j}_L(x) \frac{\bar{\hat{\rho}}_2(k)}{\sqrt{2\omega_m(k)}}\|^2_{\mathcal{B}(L^2)} \right]^{\frac{1}{2}}\\
 & & \qquad \qquad \qquad \qquad \qquad \times \left[ \int_{|x|>\frac{L}{2}}dx\, \int_{\R^n}dk\, 
  \| \one \otimes a(x,k) \chi_{\Delta}(H_m)\psi\|^2 \right]^{\frac{1}{2}}\\
 & \leq & C(\alpha, \Delta) e^{-\alpha L} \| \phi \| \times \| (\one \otimes N)^{\frac{1}{2}}\chi_{\Delta}(H_m)\psi \|. 
\end{eqnarray*}
The result then follows as in the discrete case. \endproof

\subsubsection{Removing the cutoff}\label{sssec:cont-remove}

\begin{prop}\label{prop:st-res-mass} $H_m(L)$ converges to $H_m$ in the strong resolvent sens.
\end{prop}

\pf As for Proposition \ref{prop:st-res-disc}, it suffices to show that $H_m(L)$ converges strongly to $H_m$. Let $\psi \in D(H_m^0),$
\begin{eqnarray*}
 & & \| H_m \psi -H_m(L)\psi \|^2_{\mathcal{H}}\\
 & = & \| W_m \psi -W_m(L)\psi \|^2_{\mathcal{H}} \\
 & = & \int_{|q|>\frac{L}{2}-R_1} dq\, \Big\|  \Big( \int_{|x|>\frac{L}{2}}dx\, \int_{\R^n}dk\, \rho_1(x-q)\bar{j}_L(x) 
  \frac{\hat{\rho}_2(k)}{\sqrt{2\omega_m(k)}} a^*(x,k)\\
 & &   + \int_{\R^d} dq\, \int_{|x|>\frac{L}{2}}dx\, \int_{\R^n}dk\,
  \rho_1(x-q)\bar{j}_L(x) \frac{\bar{\hat{\rho}}_2(k)}{\sqrt{2\omega_m(k)}} a(x,k)\Big) \psi(q)\Big\|^2_{\mathcal{F}}.
\end{eqnarray*}
With similar computations as the ones of Proposition \ref{prop:st-res-disc}, we get
$$
\| H_m \psi -H_m(L)\psi \|^2_{\mathcal{H}} \leq C \int_{|q|>\frac{L}{2}-R_1} dq\, \|N^{\frac{1}{2}}\psi
(q)\|^2_{\mathcal{F}} +\| \psi(q)\|^2_{\mathcal{F}}.
$$
But $N^{\frac{1}{2}}\psi(q)$ and $\psi(q)$ belong to $L^2(\R^d,\mathcal{F}),$ so the right-hand side tends to zero as $L$ goes to infinity. \endproof

\begin{prop}\label{prop:vp-mass} $E_m(L)$ converges to $E_m$ as $L$ goes to infinity.
\end{prop}

\pf Remember that $\phi_m(L)$ is a ground state of $H_m(L).$ We have
\begin{eqnarray*}
E_m & \leq & \langle \phi_m(L); H_m \phi_m(L) \rangle \\
 & \leq & E_m(L) +\langle \phi_m(L); (W_m-W_m(L)) \phi_m(L) \rangle \\
 & \leq & E_m(L) + 2 \mathcal{R}e \big( \langle \phi_m(L); \int_{|x|>\frac{L}{2}}dx\, \int_{\R^n}dk\, \rho_1(x-Q)
  \bar{j}_L(x) \\
 & & \qquad \qquad \qquad \qquad \qquad \qquad \qquad \times \frac{\hat{\rho}_2(k)}{\sqrt{2\omega_m(k)}}\otimes a(x,k)
  \phi_m(L) \rangle \big)\\
 & \leq & E_m(L) + 2 \mathcal{R}e \big( \langle e^{\alpha |Q|}\otimes \one \phi_m(L); \int_{|x|>\frac{L}{2}}dx\,
 \int_{\R^n}dk\, e^{-\alpha |Q|}\rho_1(x-Q)  \\
 & & \qquad \qquad \qquad \qquad \qquad \qquad \qquad \times \bar{j}_L(x)\frac{\hat{\rho}_2(k)}{\sqrt{2\omega_m(k)}}\otimes a(x,k)
  \phi_m(L) \rangle \big). 
\end{eqnarray*}
Then, the same computation as in Proposition \ref{prop:norm-cvg-mass} leads to
\begin{eqnarray*}
E_m & \leq & E_m(L) + K(\alpha) e^{-\frac{\alpha L}{2}} \langle \phi_m(L); \one \otimes N(|x|>\frac{L}{2}) \phi_m(L)\rangle\\
 & \leq & E_m(L) + C(\alpha) e^{-\alpha L}.
\end{eqnarray*}
So, the function $E_m(L)$ is bounded from below (and from above by $E_p^0$). Then there exists a sequence $L_n$ and $E_{\infty}$ such that 
$$
\lim_{n\to +\infty} E_m(L_n)= E_{\infty}\geq E_m.
$$ 
Now, $H_m(L_n)$ converges to $H_m$ in the strong resolvent sens and $E_m \in \sigma(H_m)$, so, for all $n,$ there exists $E(L_n)\in \sigma(H_m(L_n))$ such that 
$$
\lim_{n\to +\infty} E(L_n)= E_m.
$$
But $E(L_n)$ is bigger than $E_m(L_n)$ for all $n,$ so finally $E_m=E_{\infty}.$ The function $E_m(L)$ is then bounded with only one accumulating point $E_m$, which proves that the function converges to this point.\endproof

\noindent{\bf Proof of Theorem \ref{thm:massif}~:\ } The proof is identical to the one of Theorem \ref{thm:disc}.

\begin{rem} 
Another way to prove our results concerning the massive case would be to use the ideas of \cite{DG}-\cite{GLL}. The idea is to prove that $E_m$ is not in the essential spectrum using the Weyl criterion. For that purpose, one proves that, given a normed sequence $\psi_j$ tending weakly to zero, 
\begin{equation}\label{eq:weyl}
\liminf_{j\to \infty} \langle \psi_j;(H_m-E_m)\psi_j \rangle>0.
\end{equation}
The philosophy is that, if $\psi_j$ tends weakly to zero, it must ``escape to infinity'' in some way. In our model, if it escapes in the particle part, with the number of bosons or with their momentum in the $y$ direction (that is when $k$ tends to infinity), then the energy grows necessarily and (\ref{eq:weyl}) is certainly satisfied. Now, if it escapes with far away bosons, either in ``space'' (that is in the $x$ or $y$ direction) or in ``momentum in the $x$ direction'', the idea is that those bosons do not interact with the particle and so each of them has an energy at least $m.$ A Weyl sequence can then exist only for $E\geq E_m+m.$ A precise writing of such a proof would imply a control on the momentum of the bosons in the $x$ direction, which is the new element of our model. In our proof, such a control already exists but appears in a hidden way in Proposition \ref{prop:norm-cvg-disc}. Finally, we would like to emphasize that writing a proof using this other method would not be much shorter.
\end{rem}


\section{Proof of the main results}\label{sec:nonmassif}

The goal of this section is to prove the results of Sect. \ref{sec:results}. We start with Theorem 
\ref{thm:gdstate}. We adapt the method of \cite{G}. We will insist on the differences with this paper. The 
idea is to approach (in a way which has to be made precise) $H$ with Hamiltonians for which we know that 
they have a ground state and then to obtain the same result for $H.$ More precisely, we will use the 
following lemma:
\begin{lem}\label{lem:ah} (\cite{AH}, Lemma 4.9) Let $H, H_n (n\in \N)$ be selfadjoint operators on a Hilbert space $\mathcal{H}.$ We suppose that
\begin{itemize} 
\item[(i)] $\forall n \in \N, H_n$ has a ground state $\psi_n$ with ground state energy $E_n,$
\item[(ii)] $H_n$ tends to $H$ in the strong resolvent sens,
\item[(iii)] $\lim_{n \to +\infty} E_n=E,$
\item[(iv)] \emph{w}$-\lim_{n\to +\infty} \psi_n=\psi \neq 0.$
\end{itemize}
Then $\psi$ is a ground state of $H$ with ground state energy $E.$
\end{lem}

\subsection{Infrared cutoff}\label{ssec:ircutoff}

We denote by $\chi_{\sigma\leq\omega(k)}$ the caracteristic function of the set $\{k\in \R^n |\sigma\leq\omega(k)\}.$ For any $\sigma>0,$ we then define
\begin{eqnarray}\label{hamq:cutoff}
H^{\sigma} & := & H_0+ \int_{\R^d}dx \int_{\R^n}dk\, \rho_1(x-Q)\frac{\hat{\rho}_2(k)}{\sqrt{2\omega(k)}}\chi_{\sigma\leq\omega(k)}(k)\otimes a^*(x,k)\nonumber \\
 & & \qquad \qquad \qquad \qquad \qquad +\rho_1(x-Q)^*\frac{\bar{\hat{\rho}}_2(k)}{\sqrt{2\omega(k)}}\chi_{\sigma\leq\omega(k)}(k)\otimes a(x,k) \nonumber\\
 & = & H_0+ H_{I,\sigma}, 
\end{eqnarray}
where $H_0$ is the free Hamiltonian defined in (\ref{hamq:libre}). We want to use Lemma \ref{lem:ah} with $H$ and $H^{\sigma_n}$ where $\sigma_n$ is some sequence going to zero.

We consider a function $\tilde{\omega}_{\sigma}(k)$ satisfying
$$
\left\{ \begin{array}{l} \nabla \tilde{\omega}_{\sigma} \in L^{\infty}(\R^n), \\ 
\tilde{\omega}_{\sigma}(k)= \omega(k) \quad \textrm{si} \quad \omega(k)\geq \sigma ,\\ \inf
\tilde{\omega}_{\sigma}(k)\geq \frac{\sigma}{2}>0, \end{array} \right.
$$
and we define 
\begin{equation}
\tilde{H}^{\sigma}= H_p\otimes \one + \one \otimes \mbox{d}\Gamma(\tilde{\omega}_{\sigma}) + H_{I,\sigma}.
\end{equation}
Then we have the following result:
\begin{prop}\label{prop:cutoff} For any $\sigma>0, H^{\sigma}$ has a ground state $\psi_{\sigma}.$ We denote
by $E_{\sigma}$ its ground state energy.
\end{prop}

\noindent To prove this result we use the following lemma:
\begin{lem} (\cite{G}, Lemma 3.2) $H^{\sigma}$ has a ground state if and only if $\tilde{H}^{\sigma}$ has one.
\end{lem}

\noindent{\bf Proof of Proposition \ref{prop:cutoff}~:\ } According to the previous lemma, it suffices to show that $\tilde{H}^{\sigma}$ has a ground state. But $\tilde{H}^{\sigma}$ is a Hamiltonian of the form studied in Sect. \ref{ssec:massif}, so, according to Theorem \ref{thm:massif}, it has a ground state.\endproof

\begin{prop}\label{prop:st-res} $H^{\sigma}$ tends to $H$ in the norm resolvent sens.
\end{prop}

\pf We use Lemma A.2 of \cite{G} which says that it suffices to show that $Q^{\sigma}$ converges to $Q$ in the topology of $\mathcal{D}(Q),$ where $Q^{\sigma}$ and $Q$ are the quadratic forms associated to $H^{\sigma}$ and $H.$ But, with a similar computation to the one of Lemma \ref{lem:a-estimates}, one has
\begin{eqnarray*}
|Q(u,v)-Q^{\sigma}(u,v)| & \leq & \left( \int_{\R^d} dx\, \int_{\omega(k)\leq \sigma} dk\,
\frac{\rho_1(x-q)^2|\hat{\rho}_2(k)|^2}{2\omega^2(k)}\right)^{\frac{1}{2}} \\
 & & \qquad \qquad \qquad \times (Q(u,u)\|v\|+Q(v,v)\|u\|).
\end{eqnarray*}

\begin{cor}\label{cor:vp-cvg} $\lim_{\sigma \to 0} E_{\sigma}=E_0.$
\end{cor}

\begin{rem} As in the massive case, one has $E_{\sigma}\leq E_p^0$ for all $\sigma>0.$
\end{rem}

Using Propositions \ref{prop:cutoff} and \ref{prop:st-res} together with Corollary \ref{cor:vp-cvg}, one can see that the operators $H^{\sigma}$ and $H$ satisfy assumptions $(i)-(ii)-(iii)$ of Lemma \ref{lem:ah}. So, it remains to check condition $(iv)$ and Theorem \ref{thm:gdstate} will be proven.


\subsection{Uniform estimates}\label{ssec:unif}

\begin{lem}\label{lem:nrj-bound} There exists $C_1>0$ such that for all $\sigma>0,$
$$
\langle \psi_{\sigma}; H_0 \psi_{\sigma}\rangle \leq C_1.
$$
\end{lem}
This inequality comes from the fact that $H_{I,\sigma}$ is relatively $H_0$ bounded with infinitesimal bound, uniformly with respect to $\sigma>0.$ Of course, we need an estimate on the number of soft bosons, estimate which uses the infrared condition (IR).

\begin{lem}\label{lem:nb-bound} There exists $C_2>0$ such that for all $\sigma>0,$
$$
\langle \psi_{\sigma}; \one \otimes N \psi_{\sigma}\rangle \leq C_2.
$$
\end{lem}

\pf As in Lemma \ref{lem:pullthrough}, one can show that 
\begin{equation}\label{eq:pullthr2}
\|\one \otimes a(x,k) \psi_{\sigma}\| \leq \frac{1}{\omega(k)}\| \rho_1(x-Q)\frac{\hat{\rho}_2(k)}{\sqrt{2\omega_m(k)}}
\chi_{\omega(k)\geq \sigma}(k)\otimes \one \psi_{\sigma} \|.
\end{equation}
Thus,
\begin{eqnarray*}
\langle \psi_{\sigma}; \one \otimes N \psi_{\sigma}\rangle & = & \int_{\R^d}dx\, \int_{\R^n} dk\, \| \one \otimes a(x,k) 
  \psi_{\sigma}\|^2_{\mathcal{H}} \\
 & \leq & \int_{\R^d}dx\, \int_{\omega(k)\geq \sigma} dk\, \frac{1}{\omega^2(k)} \| \rho_1(x-Q)
  \frac{\hat{\rho}_2(k)}{\sqrt{2\omega(k)}}\otimes \one \psi_{\sigma}\|^2_{\mathcal{H}} \\
 & \leq & \int_{\R^d} dq \int_{\R^d}dx\, \int_{\omega(k)\geq \sigma} dk\, \frac{|\hat{\rho}_2(k)|^2}{2\omega^3(k)}
 |\rho_1(x-q)|^2 \|\psi_{\sigma}(q)\|^2_{\mathcal{F}}\\
 & \leq & \| \rho_1\|^2_2 \left( \int_{\R^n} dk\, \frac{|\hat{\rho}_2(k)|^2}{2\omega^3(k)} \right)
  \int_{\R^d} dq\, \|\psi_{\sigma}(q)\|^2_{\mathcal{F}}
 \leq  C_2.
\end{eqnarray*}
\endproof

We have obtained a control on the total number of bosons. However, we will also need some control (uniform with respect to $\sigma$) on the number of ``far away bosons'', that is on the following quantities: 
$\langle \psi_{\sigma}; N(|x|>R) \psi_{\sigma} \rangle$, $\langle
\psi_{\sigma}; N(|y|>S) \psi_{\sigma} \rangle$ and $\langle \psi_{\sigma}; N(|p|>P) \psi_{\sigma} \rangle$ where
$$
N(|x|>R)=\int_{|x|>R}dx\, \int_{\R^n}dk\, a^*(x,k)a(x,k),
$$
$$
N(|y|>S)=\int_{|y|>S}dx\, \int_{\R^n}dy\, \tilde{a}^*(x,y)\tilde{a}(x,y),
$$
$$
N(|p|>P)=\int_{|p|>P}dp\, \int_{\R^n}dk\, \hat{a}^*(p,k)\hat{a}(p,k).
$$
The operators $\tilde{a}$ and $\tilde{a}^*$ come from $a$ and $a^*$ via a partial Fourier transform in the $k$ variable, and the operators $\hat{a}$ and $\hat{a}^*$ via a partial Fourier transform in the $x$ variable. We then prove a result similar to Proposition \ref{prop:photdecr}:
\begin{lem}\label{lem:nb-decay1} For any $\alpha>0,$ there exists $C(\alpha)>0$ such that 
$$
\langle \psi_{\sigma}; \one \otimes N(|x|>R) \psi_{\sigma}\rangle \leq C(\alpha) e^{-\alpha R}.
$$
\end{lem}
The proof of this lemma is exactly the same to the one of Proposition \ref{prop:photdecr}. This lemma gives us a control on the number of ``far away'' bosons in the $x$ direction. Similarly one can control the number of bosons whose momentum in the $x$ direction is large:
\begin{lem}\label{lem:nb-decay2} For any $s>0,$ there exists $C(s)>0$ such that 
$$
\langle \psi_{\sigma}; \one \otimes N(|p|>P) \psi_{\sigma}\rangle \leq \frac{C(s)}{1+P^s}.
$$
\end{lem}

\pf Using (\ref{eq:pullthr2}) and a computation similar to the one of Proposition \ref{prop:photdecr}, one gets
$$
\langle \psi_{\sigma}; \one \otimes N(|p|>P) \psi_{\sigma}\rangle \leq \left( \int dk\,
\frac{|\hat{\rho}_2(k)|^2}{2\omega(k)^3} \right)\times \left( \int_{|p|>P} dp |\hat{\rho}_1(p)|^2\right),
$$
and the result follows .\endproof

Finally, to control $N(|y|>S),$ we use the following result noting that $\mbox{d}\Gamma(1-F_S(y))\leq N(|y|>\frac{S}{2}).$
\begin{lem}\label{lem:y-decay}  Let $F\in C^{\infty}_0(\R^n)$ such that 
$$
0 \leq F(y) \leq 1, \quad F(y)=1 \ \ \rm{for} \ |y|\leq 1/2, \quad \rm{and} \quad F(y)=0 \ \ \rm{for} \ 
|y|\geq 1.
$$
Let $F_S(y)=F(\frac{|y|}{S}).$ Then
$$
\lim_{\sigma\to 0, S\to +\infty} \langle \psi_{\sigma}; \textrm{d}\Gamma(1-F_S(y))\psi_{\sigma} \rangle=0.
$$
\end{lem}

\pf There is a similar result in \cite{G} (Lemma 4.5), and we essentially follow its proof. The main difference is that the norm of $\rho_1(x-Q)$ as an operator on $L^2(\R^d)$ does not depend on $x$ and is therefore not square integrable with respect to this variable. As in Sect. \ref{sssec:expbound}, to control this problem, we will use the exponential decay of the spectral projectors in the $Q$ variable (Proposition 
\ref{prop:exp-decr}).
First, one easily sees that
\begin{equation}\label{dgamma-repr} 
\mbox{d}\Gamma(1-F_S(y))=\int dx\, dk\, a^*(x,k)(1-F(\frac{|D_k|}{S}))a(x,k).
\end{equation}
We recall that for any $\sigma$ one has
$$
a(x,k)\psi_{\sigma}= (E_{\sigma}-H^{\sigma}-\omega(k))^{-1} \frac{\rho_1(x-Q)\hat{\rho}_2(k)}{\sqrt{2\omega(k)}}
\chi_{\sigma\leq\omega(k)}(k)\psi_{\sigma}.
$$
Then one can prove (\cite{G}, Prop 4.4) that 
$$
\lim_{\sigma \to 0} a(x,k)\psi_{\sigma}- (E_0-H-\omega(k))^{-1}
\frac{\rho_1(x-Q)\hat{\rho}_2(k)}{\sqrt{2\omega(k)}}\psi_{\sigma}=0
$$
in $L^2(\R^{d+n},dx\, dk;\mathcal{H}).$ Using this together with (\ref{dgamma-repr}), we then have
$$
\begin{array}{l}
\langle \psi_{\sigma}; \mbox{d}\Gamma(1-F_S(y))\psi_{\sigma} \rangle \\
\qquad = \int dx\, dk\, \langle (E_0-H-\omega(k))^{-1}\frac{\rho_1(x-Q)\hat{\rho}_2(k)}{\sqrt{2\omega(k)}}\psi_{\sigma}; \\
\qquad \qquad \qquad \qquad (1-F(\frac{|D_k|}{S}))(E_0-H-\omega(k))^{-1}\frac{\rho_1(x-Q)\hat{\rho}_2(k)}{\sqrt{2\omega(k)}}
 \psi_{\sigma}\rangle + o(\sigma^0)\\
\qquad \leq \|(E_0-H-\omega(k))^{-1}\frac{\rho_1(x-Q)\hat{\rho}_2(k)}{\sqrt{2\omega(k)}}\psi_{\sigma}\|_{L^2(\R^{d+n};\mathcal{H})} \\
\qquad \qquad \times \|(1-F(\frac{|D_k|}{S}))(E_0-H-\omega(k))^{-1}\frac{\rho_1(x-Q)\hat{\rho}_2(k)}{\sqrt{2\omega(k)}}
 \psi_{\sigma}\|_{L^2(\R^{d+n};\mathcal{H})}+ o(\sigma^0)\\
\qquad \leq \|(E_0-H-\omega(k))^{-1}\frac{\rho_1(x-Q)e^{-\alpha|Q|}\hat{\rho}_2(k)}{\sqrt{2\omega(k)}}\|_{L^2(\R^{d+n};\mathcal{B}(\mathcal{H}))}
  \times \|e^{\alpha|Q|} \psi_{\sigma}\|_{\mathcal{H}}\\
\qquad \qquad \times \|(1-F(\frac{|D_k|}{S}))(E_0-H-\omega(k))^{-1} \frac{\rho_1(x-Q)e^{-\alpha|Q|}\hat{\rho}_2(k)}{\sqrt{2\omega(k)}}\|_{L^2(\R^{d+n};\mathcal{B}(\mathcal{H}))}\\
\qquad \qquad \qquad \times \|e^{\alpha|Q|} \psi_{\sigma}\|_{\mathcal{H}}+ o(\sigma^0).\\
\end{array}
$$
We check that $(E_0-H-\omega(k))^{-1}\frac{\rho_1(x-Q)e^{-\alpha|Q|}\hat{\rho}_2(k)}{\sqrt{2\omega(k)}}$ belongs to $L^2(\R^{d+n};\mathcal{B}(\mathcal{H})),$ using the fact that $\|(E_0-H-\omega(k))^{-1}\|\leq \omega(k)^{-1}$ and condition (IR). Thus
$$
\lim_{S\to +\infty} \|(1-F(\frac{|D_k|}{S}))(E_0-H-\omega(k))^{-1}
\frac{\rho_1(x-Q)e^{-\alpha|Q|}\hat{\rho}_2(k)}{\sqrt{2\omega(k)}}\|_{L^2(\R^{d+n};\mathcal{B}(\mathcal{H}))}=0.
$$
Moreover $\|e^{\alpha|Q|} \psi_{\sigma}\|_{\mathcal{H}}$ is uniformly bounded (w.r.t $\sigma$), which can be proven as for $\|e^{\alpha|Q|} \psi_m(L)\|_{\mathcal{H}}$ (see Sect. \ref{ssec:massif}), and the result follows.\endproof


\subsection{Proof of Theorem \ref{thm:gdstate}}\label{ssec:proof}

We have seen that the only thing we had to check was condition $(iv)$ of Lemma \ref{lem:ah}. The unit ball of $\mathcal{H}$ is weakly compact, so there exists a sequence $\sigma_n \to 0$ and $\psi \in \mathcal{H}$ such that $\psi_{\sigma_n}$ converges weakly to $\psi.$ It then suffices to prove that $\psi\neq 0.$ The idea is to find a compact operator $K$ such that for any $n$ large enough one has such an estimate:
\begin{equation}\label{eq:unif-est}
\|K \psi_{\sigma_n}\| \geq \delta>0.
\end{equation}
This will ensure that $\psi$ is non zero. Indeed, $K$ is compact, so $K \psi_{\sigma_n}$ tends strongly to $K\psi.$ If $\psi$ was zero then $\|K \psi_{\sigma_n}\|$ would go to zero, which enters in contradiction with (\ref{eq:unif-est}).

Let us then take $F\in C^{\infty}_0(\R^n)$ and $G\in C^{\infty}_0(\R^d)$ satisfying the conditions of Lemma \ref{lem:y-decay}. Remembering that $p$ is the variable conjugate to $x,$ \emph{i.e. } $p=-i\nabla_x$ on $L^2(\R^{d+n}, dx\, dk),$ one has the following inequalities:
\begin{equation}\label{eq:gammaf}
(1-\Gamma(F_S(y)))^2 \leq (1-\Gamma(F_S(y))) \leq \mbox{d}\Gamma(1-F_S(y)),
\end{equation}
\begin{equation}\label{eq:gammag1}
(1-\Gamma(G_R(x)))^2 \leq (1-\Gamma(G_R(x))) \leq \mbox{d}\Gamma(1-G_R(x)) \leq N(|x|>\frac{R}{2}),
\end{equation}
\begin{equation}\label{eq:gammag2}
(1-\Gamma(G_P(p)))^2 \leq (1-\Gamma(G_P(p))) \leq \mbox{d}\Gamma(1-G_P(p)) \leq N(|p|>\frac{P}{2}).
\end{equation}
Finally, let $\chi(s\leq s_0)$ be a function with support in $\{ |s|\leq s_0\}$ and equal to $1$ in $\{ |s|\leq \frac{s_0}{2}\}.$ For any non negative $\theta, P, R$ and $S,$ we define
\begin{equation}\label{def:compact}
K(\theta,P,R,S):= \chi(N\leq \theta)\chi(H_0\leq \theta)\Gamma(F_S(y))\Gamma(G_R(x))\Gamma(G_P(p)).
\end{equation}
The assumptions on $F, G, \chi$ and $\omega$ ensure that $K(\theta,P,R,S)$ is compact for any $\theta, P, R$ and $S.$

Using Lemmas \ref{lem:nrj-bound} and \ref{lem:nb-bound}, there exists $\theta_0>0$ such that, for all $n,$ one has: 
\begin{equation}
\|(1-\chi(N\leq \theta))\psi_{\sigma_n}\|\leq \frac{1}{10}, \|(1-\chi(H_0\leq \theta))\psi_{\sigma_n}\|\leq \frac{1}{10}.
\end{equation} 
Likewise, using Lemmas \ref{lem:nb-decay1} and \ref{lem:nb-decay2} together with inequalities (\ref{eq:gammag1}) and (\ref{eq:gammag2}), there exist $R_0,P_0>0$ such that, for all $n,$ one has: 
\begin{equation}
\|(1-\Gamma(G_R(x)))\psi_{\sigma_n}\| \leq \frac{1}{10}, \|(1-\Gamma(G_P(p)))\psi_{\sigma_n}\| \leq \frac{1}{10}.
\end{equation}
Finally, using Lemma \ref{lem:y-decay} and (\ref{eq:gammaf}), there exist $S_0>0$ and $n_0$ such that, for all $n\geq n_0,$ one has:
\begin{equation}
\|(1-\Gamma(F_S(y)))\psi_{\sigma_n}\| \leq \frac{1}{10}.
\end{equation}
Then, for any $n\geq n_0:$ 
\begin{eqnarray*}
\|\psi_{\sigma_n}\| & \leq & \|(1-\chi(N\leq \theta_0))\psi_{\sigma_n}\|+\|\chi(N\leq \theta_0)(1-\chi(H_0\leq
  \theta_0))\psi_{\sigma_n}\| \\
 & & + \|\chi(N\leq \theta_0)\chi(H_0\leq \theta_0)(1-\Gamma(G_{R_0}(x)))\psi_{\sigma_n}\| \\
 & & + \|\chi(N\leq \theta_0)\chi(H_0\leq \theta_0)\Gamma(G_{R_0}(x))(1-\Gamma(G_{P_0}(p)))\psi_{\sigma_n}\| \\
 & & + \|\chi(N\leq \theta_0)\chi(H_0\leq \theta_0)\Gamma(G_{R_0}(x))\Gamma(G_{P_0}(p))(1-\Gamma(F_{S_0}(y)))\psi_{\sigma_n}\| \\
 & & + \|K(\theta_0,P_0,R_0,S_0)\psi_{\sigma_n}\| \\
 & \leq & \frac{1}{2} +\|K(\theta_0,P_0,R_0,S_0)\psi_{\sigma_n}\|. 
\end{eqnarray*}
But $\|\psi_{\sigma_n}\|=1$ for all $n,$ thus 
$$
\|K(\theta_0,P_0,R_0,S_0)\psi_{\sigma_n}\| \geq \frac{1}{2},
$$
for any $n\geq n_0,$ which is an estimate of the form (\ref{eq:unif-est}).\endproof


\subsection{Proof of Proposition \ref{prop:noground}} \label{ssec:noground}

The idea of the proof is adapted from \cite{DG2}. Once again, one of the main tools is the pullthrough 
formula, which comes from the commutator between $H$ and annihilation operators 
\begin{equation}\label{eq:pull}
[H,\one \otimes a(x,k)]=-\omega(k) \one \otimes a(x,k)-\rho_1(x-Q)\frac{\hat{\rho}_2(k)}{\sqrt{2\omega(k)}}
\otimes \one.
\end{equation}
In order to get our result we will need to use this formula taking into account the membranes alltogether, 
which, on a formal level, means that we will integrate the previous formula over the ``$x$-space''. It is 
therefore more convenient to look at the Hamiltonian not in the $(x,k)$ variables but in the $(p,k)$ 
variables, where $p$ is the variable conjugate to $x$ \emph{via} Fourier transform, and then consider the 
value $p=0$. In such variables, the pullthrough formula just becomes  
\begin{equation}\label{eq:pull2}
[H,\one \otimes \hat{a}(p,k)]=-\omega(k) \one \otimes \hat{a}(p,k)-\hat{\rho}_1(p)e^{-ipQ}
\frac{\hat{\rho}_2(k)}{\sqrt{2\omega(k)}}\otimes \one.
\end{equation}

Suppose now that $\Psi\in \mathcal{H}$ satisfies $H\Psi=E_0\Psi,$ where $E_0$ is the ground state energy of $H.$ We will show that $\Psi=0.$
We apply equation (\ref{eq:pull2}) on such a vector. One then gets the following equality
$$
\one \otimes \hat{a}(p,k)\ \Psi=-(H+\omega(k)-E_0)^{-1}\left( \frac{\hat{\rho}_1(p)e^{-ipQ}
\hat{\rho}_2(k)}{\sqrt{2\omega(k)}}\otimes \one\right) \Psi.
$$
We denote with an exponent $(m)$ the component of a vector in the $m$-particle sector. We have, for any $m$,
$$
(\one \otimes \hat{a}(p,k)\ \Psi)^{(m)}(p_1,k_1,\dots,p_m,k_m)=\Psi^{(m+1)}(p,k,p_1,k_1,\dots,p_m,k_m)
$$
and the righthand side is square integrable with respect to all its arguments because 
$\Psi \in \mathcal{H}.$ Therefore, for all $m,$ 
$$
\Phi^{(m)}(p,k):= \left( -(H+\omega(k)-E_0)^{-1} \frac{\hat{\rho}_1(p)e^{-ipQ}
\hat{\rho}_2(k)}{\sqrt{2\omega(k)}}\otimes \one \ \Psi \right)^{(m)}
$$
is square integrable with respect to $(p,k).$ On the other hand, it is a continuous function on $\R^d\times (\R^n-\{0\}).$ Then, for any $p_0\in \R^d, \, \Phi^{(m)}(p_0,k)$ is a well defined function of $k$ and it is square integrable.  As we have said previously, we consider the value $p_0=0.$ But
$$
\Phi^{(m)}(0,k)= \frac{\hat{\rho}_1(0)\hat{\rho}_2(k)}{\sqrt{2}\omega(k)^{\frac{3}{2}}} \Psi^{(m)},
$$
which is not square integrable if the infrared condition is violated, unless $\hat{\rho}_1(0)\Psi^{(m)}=0.$
By assumption, $\hat{\rho}_1(0)\neq 0,$ so $\Psi^{(m)}=0$ for all $m$ which means that $\Psi=0.$  
\endproof


\section{A classical interpretation of the infrared problem}\label{sec:infrarouge}

In this section, we would like to say a few words about the infrared problem. We know that this 
condition is necessary and sufficient for the existence of a ground state in the case of the Nelson model, 
and sufficient and ``almost'' necessary in our model.

On the other side, in \cite{A2}, the author shows that if we consider another \emph{ad hoc} representation of the canonical commutation relations the Nelson model without infrared condition has a ground state. In some sens, this representation regularizes the infrared singularity and is of course not unitarily equivalent to the Fock one. One could think that the same approach should work in our case. However, it turns out that this is not true. To explain why, we will briefly explain the physical origin of this representation. It will allow us to see that this procedure can apply to our model but does not have the same regularising effect.

We would like to explain the idea which is behind this change of representation coming back to classical mechanics \cite{DB}. It will then allow us to show the difference between the Nelson model and ours. We thus consider a classical Hamiltonian of the form
$$
H=\frac{1}{2}\int_A d\mu(\alpha) (\omega(\alpha)^2 \phi(\alpha)^2 +\pi(\alpha)^2),
$$
where $\omega(\alpha)$ is some almost everywhere non negative function. We will simply write $X$ instead of $(\phi,\pi).$ The Hamiltonian flow can be written as  
$$
X_t=\Phi_t X_0= \cos(\omega t)X_0- \sin(\omega t) JX_0, \quad \rm{where} \ \
J= \left( \begin{array}{rr} 0 & -\omega^{-1} \\ \omega & 0 \\ \end{array} \right).
$$
One can see that $J^2=-\one.$

Now, for any $s\in \R,$ we define
$$
\mathcal{D}(\omega^s):= \{ \phi \in L^2 | \omega^s \phi \in L^2\},
$$
and $[\mathcal{D}(\omega^s)]$ its closure for the norm $\| \phi \|_s= \| \omega^s \phi \|_{L^2}.$ Given $s,r \in \R,$ we define
$$
\mathcal{H}_{s,r}:=[\mathcal{D}(\omega^s)]\times [\mathcal{D}(\omega^r)].
$$
The operator $J$ is well defined on $\mathcal{H}_{s,r}$ (as a bounded operator) if and only if $r=s-1.$ On the other side, the symplectic form 
$$
\sigma(X_1,X_2)=\int_A d\mu(\alpha)(\phi_1(\alpha) \pi_2(\alpha)-\phi_2(\alpha) \pi_1(\alpha))
$$
is meaningfull only on spaces of the form $\mathcal{H}_{s,-s}.$ So $J$ and $\sigma$ are both well defined only on
$$
\mathcal{H}=[\mathcal{D}(\omega^{\frac{1}{2}})]\times [\mathcal{D}(\omega^{-\frac{1}{2}})].
$$
On $\mathcal{H}$, we consider the following complex structure $\langle \,\, ; \,\,\rangle$ defined as
$$
\langle X_1;X_2\rangle= \sigma(X_1,JX_2)+ i \sigma(X_1,X_2).
$$
One can then identify $\mathcal{H}$ and $L^2(A, d\mu,\C)$ \emph{via} the following isometry:
$$
(\phi,\pi)\in \mathcal{H}\to \sqrt{\omega(\alpha)}\phi(\alpha)+ \frac{i}{\sqrt{\omega(\alpha)}}\pi(\alpha) \in
L^2(A, d\mu,\C).
$$
and if we define
$$
a(\alpha):=\frac{1}{\sqrt2} \left( \sqrt{\omega(\alpha)}\phi(\alpha)+ \frac{i}{\sqrt{\omega(\alpha)}}\pi(\alpha)\right) \in
L^2(A, d\mu,\C), 
$$
one can rewrite $H$ as
$$
H=\int_A \omega(\alpha) a^*(\alpha)a(\alpha) d\mu(\alpha).
$$
The Poisson bracket associated to $\sigma$ is
$
\{ \phi(\alpha),\pi(\alpha)\}= \delta_{\mu}(\alpha-\alpha'),
$
where $\delta_{\mu}$ is defined by
$$
\int_A f(\alpha') \delta_{\mu}(\alpha-\alpha')d\mu(\alpha')=f(\alpha).
$$
It is easy to see that
$
\{a(\alpha),a^*(\alpha')\}=-i \delta_{\mu}(\alpha-\alpha').
$
Those relations are the classical equivalent of the relations (\ref{ccr2}). To write the quantum version of this model, one then consider the Fock space over $L^2(A, d\mu,\C).$

Consider now the Nelson model, \emph{i.e.} $A=\R^d$ and $d\mu(\alpha)= dk.$ If we consider a particle which interacts with this field and which is on the other hand submitted to a confining potential $V$ such that $\min V=V(0),$ the equilibrium point of the system which correpsonds to the minimum of the energy can be written as $(q_*,\phi_*,p_*,\pi_*)=(0,-\frac{\hat{\rho}}{\omega^2},0,0)$ and $(\phi_*,\pi_*)$ is in $\mathcal{H}$ if and only if $\frac{\hat{\rho}}{\omega^{3/2}}\in L^2(\R^d),$ which is exactly the condition (IR). One should remind that, for this model, this condition is necessary and sufficient to have a ground state. In other words, the minimum $(q_*,\phi_*,p_*,\pi_*)$ of the classical Hamiltonian belongs to $\mathcal{H}$ if and only if (IR) is satisfied. In this way, one can say that the condition to have a ground state is the same on both classical and quantum level.

The representation considered in \cite{A2} corresponds, on the classical level, to the affine space ``$\tilde{\mathcal{H}}=\mathcal{H}+(\frac{\hat{\rho}}{\omega^2},0)$'', more precisely, one considers the following symplectic transformation:
$$
\tilde{q}=q,\tilde{\phi}=\phi+\frac{\hat{\rho}}{\omega^2},\tilde{p}=p,\tilde{\pi}=\pi.
$$
Here, $\tilde{\phi}$ represents the difference between the field and its equilibrium position. We already note that $0\notin \tilde{\mathcal{H}}$ if and only if (IR) is not satisfied. One then sees that the two equilibrium points, before ($\phi_*=0$) and after ($\phi_*=-\frac{\hat{\rho}}{\omega^2}$) having turned on the interaction with the particle, are not in the same space. This is this phenomenon which, on the quantum level, expresses that the ground state exists but in another representation, non-equivalent to the Fock one. One sometimes reads that the ``ground state is not in the Fock space'' (within the context of ``Van Hove Hamiltonians'' for example \cite{D}-\cite{Fr}-\cite{VH}).

In those new variables, the Hamiltonian of the whole system then writes
\begin{eqnarray}\label{tildeham}
\tilde{H}(\tilde{q},\tilde{\phi},\tilde{p},\tilde{\pi}) & = & \frac{1}{2} \int (\omega^2 \tilde{\phi}^2+\tilde{\pi}^2)
  +\int \hat{\rho}(k)(e^{-ikq}-1)\tilde{\phi}\nonumber \\
 & &  +\frac{\tilde{p}^2}{2}+ V(\tilde{q})\underbrace{-\int
 \frac{|\hat{\rho}(k)|^2e^{-ikq}}{\omega^2}}_{=W(\tilde{q})\,\, \textrm{bounded}}+ \underbrace{\int
 \frac{|\hat{\rho}(k)|^2}{2\omega^2}}_{=\textrm{constant}}.
\end{eqnarray}
If we want to study the system near the new equilibriun, one then has to chose the phase space such that $\tilde{\phi}=0$ belongs to it. It is then natural to study $\tilde{H}$ not on $\tilde{\mathcal{H}}$ but on $\mathcal{H}.$ In other words, to make sure that the equilibrium point for the interacting system is in the phase space, one has to consider another space. If one does so, condition (IR) is then satisfied even for $d=3$:
$$
\frac{\hat{\rho}(k)(e^{-ikq}-1)}{\omega^{\frac{3}{2}}}\in L^2(\R^d).
$$
Then, if one quantizes $\tilde{H}$, one obtains a model in which a ground state exists. This is precisely what Arai does in \cite{A2}, but without explaining it this way. However, the same transformation in our model does not make things ``better.'' Indeed, condition (IR) becomes, after the same transform: 
$$
\frac{[\rho_1(x-q)-\rho_1(x)]\hat{\rho}_2(k)}{\omega(k)^{\frac{3}{2}}} \in L^2(\R^d\times\R^n),
$$
which is still not satisfied if $n=3$ unless $\hat{\rho}_2(0)=0.$


\noindent \textbf{Acknowledgments:} Part of this work was supported by the Postdoctoral Training Program
HPRN-CT-2002-0277. The author wishes to thank S. De Bi\`evre for many enjoyable discussions and useful comments.


\end{document}